\pdfoutput=1
\documentclass[10pt,twocolumn,twoside]{IEEEtran}

\usepackage{amsmath}
\usepackage{amssymb}
\usepackage{subcaption}
\usepackage{algorithmic}
\usepackage{algorithm}
\usepackage{mathtools}
\usepackage{xcolor}
\usepackage{makecell}
\usepackage{multirow}
\usepackage{graphicx}
\usepackage{marginnote}
\usepackage{float}
\usepackage{gensymb}
\usepackage{amsthm}
\usepackage[utf8]{inputenc}
\usepackage[english]{babel}
\usepackage{cite}
\usepackage{tikz}
\usepackage{graphicx}
\usetikzlibrary{calc,decorations.pathmorphing,patterns,arrows.meta}
\newtheorem{remark}{Remark}
\newtheorem{assumption}{Assumption}

\usepackage{tikz}
\usetikzlibrary{fit,backgrounds}

\usepackage{amsmath}
\usepackage{algorithmic}

\usepackage{array}
\usepackage{endnotes}

\DeclareMathOperator{\col}{\mathrm{col}}
\DeclareMathOperator{\diag}{\mathrm{diag}}

\DeclareMathAlphabet{\doublestruck}{U}{BOONDOX-ds}{m}{n}

\newcommand{\R}[0]{\mathbb{R}}

\newcommand{\Ccal}[0]{\mathcal{C}}

\newcommand{\Ecal}[0]{\mathcal{E}}

\newcommand{\Gcal}[0]{\mathcal{G}}

\newcommand{\Ncal}[0]{\mathcal{N}}

\newcommand{\Pcal}[0]{\mathcal{P}}

\newcommand{\Ucal}[0]{\mathcal{U}}

\newcommand{\Xcal}[0]{\mathcal{X}}

\newcommand{\Wcal}[0]{\mathcal{W}}
\newcommand{\Zcal}[0]{\mathcal{Z}}

\usepackage{xstring}

\newtheorem{lemma}{Lemma}
\newtheorem{theorem}{Theorem}
\newtheorem{corollary}{Corollary}
\newtheorem{definition}{Definition}
\usepackage[hidelinks]{hyperref} 
\IEEEoverridecommandlockouts 

\begin{document}

\title{\LARGE \bf
Online Adaptive Clustering \\ in Robust Distributed Model Predictive Control}


 \author{Stefan van Dijk, W.P.M.H. Heemels, Paula Chanfreut Palacio
\thanks{This research received support from the Eindhoven Artificial Intelligence Systems  Institute (EAISI) }
\thanks{S.~van~Dijk, W.P.M.H.~Heemels and P. Chanfreut are with the Control Systems Technology Section, Department of Mechanical Engineering, Eindhoven University of Technology, The Netherlands. E-mails:
        {\tt\small  \{s.v.dijk, m.heemels, p.chanfreut.palacio\}@tue.nl}.}%
}

%
%

\maketitle

\begin{abstract}
This paper proposes a dynamic clustering framework for tube-based model predictive control (MPC) of coupled multi-agent systems. Agents within a cluster coordinate their dynamics explicitly, while couplings from agents outside the cluster are treated as bounded disturbances. For each admissible cluster, we construct a state-dependent mismatch measure from a robust upper bound on its predicted closed-loop cost and an optimistic, partition-independent lower bound. Combining this mismatch with a communication penalty enables online selection of the clustering topology, ranging from centralized to fully decentralized operation. The resulting switching controller is shown to preserve recursive feasibility and constraint satisfaction, and to render the closed-loop system practically asymptotically stable. Furthermore, the mismatch provides an upper bound on the finite-horizon performance difference with respect to centralized MPC. An eight-agent spring--damper example illustrates how the controller adapts cooperation while maintaining performance close to centralized MPC with reduced communication.
\end{abstract}

\section{Introduction}

Model predictive control (MPC) is well suited to constrained multi-agent systems, but centralized implementation requires global information and an optimization problem whose size grows with the network~\cite{Negenborn2014DMPCSurvey}. Distributed and decentralized MPC reduce these requirements by decomposing the control problem, typically at the expense of performance or robustness when dynamic couplings between agents are significant~\cite{camponogara2002distributed,venkat2006distributed}.

Cluster-based, or coalitional, control provides an intermediate architecture in which groups of agents coordinate internally while interactions between different clusters are treated separately~\cite{maestre2014coalitional,fele2017coalitional}. Changing the partition of the agent set therefore allows the control architecture to range continuously from centralized to fully decentralized operation. This creates a natural trade-off: larger clusters can account for more dynamic coupling but require more communication and larger optimization problems~\cite{ALGHERWI2010270}. Existing coalitional MPC approaches have considered topology changes explicitly, including schemes that optimize the timing of transitions between communication structures while preserving recursive feasibility and robust stability~\cite{MASERO20203342, Baldivieso2021}.

A central question is consequently how the clustering topology should be selected online~\cite{Riccardi2026}. Existing approaches employ, among others, structural or coupling-based criteria and data-driven clustering methods~\cite{zheng2018coupling,chanfreut2021survey,labella2022supervised,chanfreut2022distributed}. Recent work has also considered cost-based coalition formation, using upper bounds on coalition value to reduce the computational and information-sharing requirements of evaluating candidate coalitions~\cite{rickard2026coalition}. For constrained systems, however, changing the partition must additionally preserve feasibility and closed-loop stability. Moreover, a useful switching criterion should quantify when the additional coordination provided by a larger cluster is relevant to control performance. 

In this paper, we address these requirements using a\textit{ partition-dependent} tube MPC formulation. More precisely, for every admissible cluster, intra-cluster couplings are included in the prediction model, whereas inter-cluster couplings are treated as bounded disturbances using standard tube-MPC constructions~\cite{langson2004robust,mayne2005robust,baldivieso2021tube}. The required tubes, tightened constraints, and terminal ingredients are computed offline.

To determine the partition online, we introduce a \emph{cluster mismatch} measure. For each cluster, a robust upper bound on the predicted real cost is obtained from the tube-MPC solution, while an optimistic lower bound is obtained by allowing the neighboring coupling disturbances to take their most favorable admissible values. Their difference quantifies the uncertainty in the achievable performance associated with that cluster. Since the lower bounds are additive and partition independent, the resulting mismatch can also be related to the predicted performance of centralized MPC.

The cluster mismatch is combined with a communication penalty to assign a state-dependent value to each admissible cluster. The controller then selects the feasible partition with minimum aggregate value and applies the corresponding cluster-level tube-MPC controllers. This extends our earlier state-dependent analysis of regions in which neighboring interactions have limited performance impact~\cite{vandijk2025identifying} from individual communication links to online selection among complete clustering topologies. 
The proposed framework can be viewed as a switched system~\cite{Liberzon2003}, where switching occurs at the control layer through changes in the clustering topology, as also considered in distributed MPC~\cite{Li2020}, while the underlying physical dynamics remain unchanged.

The main contributions are:
\begin{enumerate}
    \item a partition-dependent tube-MPC formulation that represents centralized, decentralized, and intermediate clustering architectures within a common framework;
    \item a computable mismatch measure based on robust upper and optimistic lower predicted-cost bounds, together with a finite-horizon performance bound relative to centralized MPC; and
    \item an online clustering algorithm for which recursive feasibility, constraint satisfaction, and practical asymptotic stability are established under the stated tube-MPC assumptions.
\end{enumerate}

The method is demonstrated on an eight-agent spring--damper network. The numerical results illustrate state-dependent changes in cooperation, including temporary recoupling, while obtaining closed-loop costs close to centralized MPC and reducing communication when additional coordination is unnecessary.

The remainder of the paper is organized as follows. Section~\ref{sec:sys_desc} introduces the multi-agent system and admissible clustering structures. Section~\ref{sec:tube_mpc} develops the cluster-level tube-MPC formulation and the mismatch measure. Section~\ref{sec:guarantees} presents the dynamic clustering algorithm and its performance, feasibility, and stability guarantees. Section~\ref{sec:system} presents the numerical example, and Section~\ref{sec:conclusion} presents the conclusions and future work we draw from this work.

\subsubsection*{Notation}
Let $\mathbb{N}_{[a,b]}:=\{a,a+1,\ldots,b\}$ for integers $a\leq b$, and let $\mathbb{R}^n$ and $\mathbb{R}^{n\times m}$ denote the real vector and matrix spaces of corresponding dimensions. For a vector $x$, $\|x\|$ denotes the Euclidean norm. For a matrix $M$, $M^\top$ denotes its transpose, $\rho(M)$ its spectral radius, and $\lambda_{\min}(M)$ its minimum eigenvalue when $M$ is symmetric. The relations $M\succ0$ and $M\succeq0$ denote positive definiteness and positive semidefiniteness, respectively, and $\operatorname{int}(\mathcal X)$ denotes the interior of a set $\mathcal X$. For sets $\mathcal A$ and $\mathcal B$, $\mathcal A\oplus\mathcal B$ and $\mathcal A\ominus\mathcal B$ denote the Minkowski sum and Pontryagin difference, respectively, and $|\mathcal A|$ denotes the cardinality of a finite set. The Cartesian product of sets $\mathcal X_i$, $i\in\mathcal I$, is denoted by $\prod_{i\in\mathcal I}\mathcal X_i$. The operators $\operatorname{col}_{i\in\mathcal I}(x_i)$ and $\operatorname{blkdiag}_{i\in\mathcal I}(M_i)$ denote column-wise stacking and block-diagonal concatenation, respectively. For predicted quantities, we use the notation $x_{t|k}$, where $t$ is the prediction step and $k$ the current sampling instant. A superscript $\star$ denotes an optimal solution, while $\tilde{(\cdot)}$ and $\hat{(\cdot)}$ denote feasible candidate sequences constructed for evaluation and recursive-feasibility arguments, respectively.

\section{Problem Formulation}\label{sec:sys_desc}
We introduce the system dynamics and the multi-agent control network considered in this paper. We then formulate the control objectives addressed in this work.
\subsection{Dynamics}
Consider a network of $N\in \mathbb{N}$ coupled agents with linear, coupled dynamics given by
\begin{equation}\label{eq:model}
\begin{split}
x_i(k+1) &= A_{ii} x_i(k) + B_{ii} u_i(k) + w_i(k),\\
w_i(k) &= \sum_{j \in \mathcal{N}_i} \underbrace{(A_{ij} x_j(k)+B_{ij} u_j(k))}_{=:w_{ij}(k)},
\end{split}
\end{equation}
for all $i \in \mathbb{N}_{[1,N]} := \{1, \dots, N\}$, at discrete time $k\in\mathbb{Z}$. Here, $x_i(k) \in \mathbb{R}^{n_i}$ and $u_i(k) \in \mathbb{R}^{m_i}$ denote the state and control input of agent $i$, and $w_i(k) \in \mathbb{R}^{n_i}$ captures the aggregated coupling effects of neighboring agents. The neighbor set of agent $i$ is defined as
\begin{equation}
\mathcal{N}_i := \{ j \in \mathbb{N}_{[1,N]} \setminus \{i\} \ : \ [A_{ij} \ B_{ij}] \neq \mathbf{0} \},
\end{equation}
which contains all agents with direct dynamic influence on $i$. Each agent is subject to local, polyhedral state and input constraints of the form
\begin{equation}\label{eq:constraints_local}
\begin{split}
\mathcal{X}_i &:= \{ x \in \mathbb{R}^{n_i} \;|\; C_{\mathrm{x},i} x \leq \mathbf{1} \}, \\
\mathcal{U}_i &:= \{ u \in \mathbb{R}^{m_i} \;|\; C_{\mathrm{u},i} u \leq \mathbf{1} \},
\end{split}
\end{equation}
where $C_{\mathrm{x},i}$ and $C_{\mathrm{u},i}$ are matrices of appropriate dimension, that define convex, compact polytopic sets, and $\mathbf{1}$ denotes a vector of ones of corresponding size. The coupling term $w_i(k)$ is likewise bounded within a compact convex set
\begin{equation}\label{eq:wi_bounds}
\mathcal{W}_i := \bigoplus_{j \in \mathcal{N}_i} \left( A_{ij} \mathcal{X}_j \oplus B_{ij} \mathcal{U}_j \right),
\end{equation}
where $\oplus$ denotes the Minkowski sum. The stage performance cost of each agent $i \in \mathbb{N}_{[1,N]}$ is given by
\(
\ell_i(\chi_i, \nu_i) := \chi_i^\top Q_i \chi_i + \nu_i^\top R_i \nu_i,
\)
where $Q_i\succ 0$ and $R_i\succ 0$ are the state and input weighting
matrices, respectively. This results in the global stage cost, with arguments $\chi_i\in \R^{n_i}, \nu_i\in \R^{m_i}$,
\begin{equation}
    \ell(\chi,\nu) = \sum_{i=1}^N \ell_i(\chi_i,\nu_i), \qquad \chi = [\chi_i]_{i=1}^N, \qquad \nu = [\nu_i]_{i=1}^N. 
\end{equation}
Hence, interactions between agents arise through the coupled dynamics and are reflected only implicitly in the constraints and objective function.

\subsection{Multi-Cluster Network}
The graph structure of the system dynamics $\mathcal{G}_{\mathrm{graph}} = (\mathcal{N},\mathcal{E})$, where the node set $\mathcal{N} := \mathbb{N}_{[1,N]}$ represents the collection of agents, and the set of directed edges is given by
\begin{equation}
    \left((i,j)\in \mathcal{E} \iff i \in \mathcal{N}_j\right)
\end{equation}
Hence, an edge exists between agent $i$ and $j$, if the state and input of agent $j$ affects the dynamics of agent $i$. We proceed with considering only undirected graphs, which means $i\in \Ncal_j \iff j \in \Ncal_i$ The graph thus encodes the dynamical coupling structure of the overall multi-agent system. 

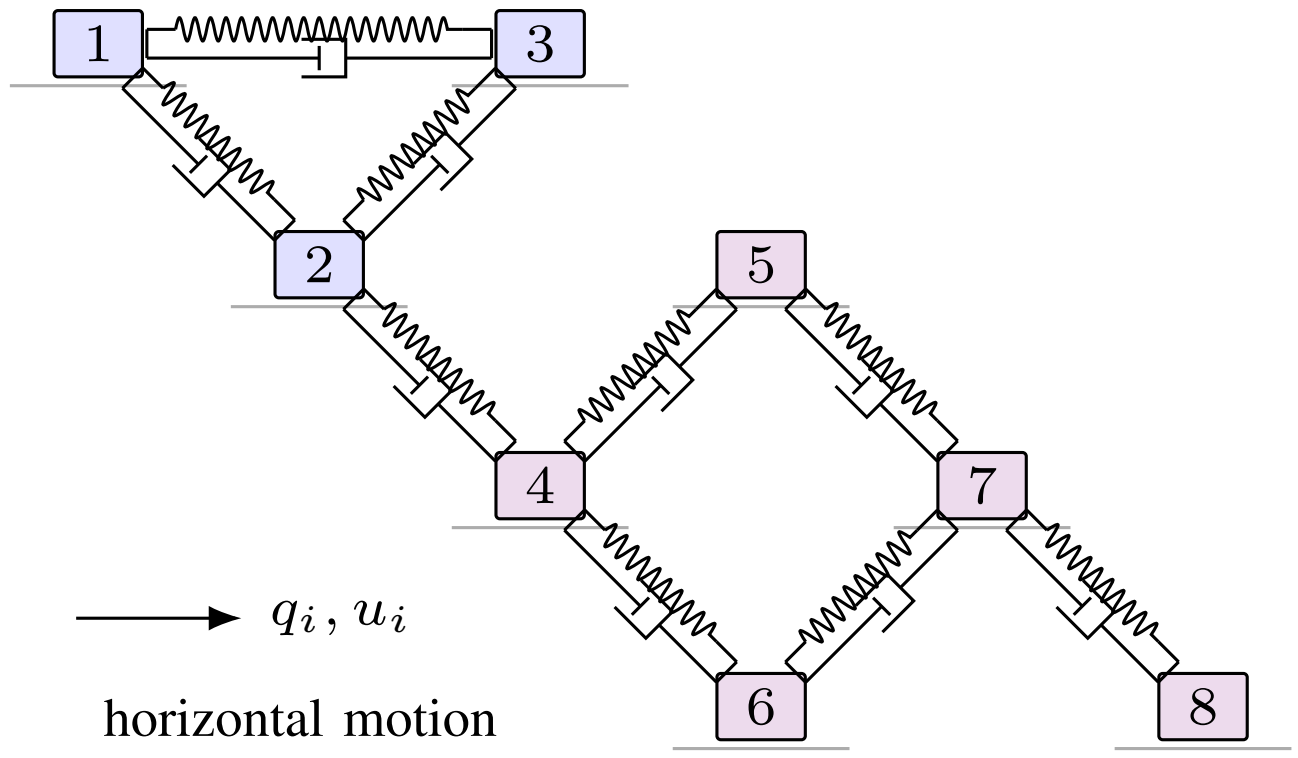
\begin{figure}[t]
\centering
\resizebox{\columnwidth}{!}{\begingroup
\newcommand{\MPCrailLink}[2]{%
 \pgfmathanglebetweenpoints{\pgfpointanchor{#1}{center}}{\pgfpointanchor{#2}{center}}%
 \let\MPCangle\pgfmathresult
 \path let \p1=($(#2.center)-(#1.center)$) in
 \pgfextra{\pgfmathsetmacro{\MPCtmp}{veclen(\x1,\y1)/1cm}\xdef\MPClength{\MPCtmp}};
 \begin{scope}[shift={(#1.center)},rotate=\MPCangle]
 \pgfmathsetmacro{\L}{\MPClength-0.22}
 \draw (0.22,-.065)--(0.22,.065);
 \draw (\L,-.065)--(\L,.065);
 \draw (0.22,.065)--(.35,.065);
 \draw[decorate,decoration={coil,aspect=0,segment length=2pt,amplitude=1.5pt}]
 (.35,.065)--(\L-.13,.065);
 \draw (\L-.13,.065)--(\L,.065);
 \draw (0.22,-.065)--(.5*\MPClength,-.065);
 \draw (.5*\MPClength,-.12)--(.5*\MPClength,-.01);
 \draw (.5*\MPClength-.08,-.15)--(.5*\MPClength+.12,-.15)
 --(.5*\MPClength+.12,.02)--(.5*\MPClength-.08,.02);
 \draw (.5*\MPClength+.12,-.065)--(\L,-.065);
 \end{scope}
}
\begin{tikzpicture}[line width=.4pt,
 mass/.style={draw,rounded corners=.5pt,minimum width=4mm,
 minimum height=3mm,inner sep=0pt,font=\scriptsize}]
\foreach \i/\x/\y in {1/0/1,2/1/0,3/2/1,4/2/-1,5/3/0,6/3/-2,7/4/-1,8/5/-2}{
 \coordinate (p\i) at (\x,\y);
 \draw[gray!65] (\x-.40,\y-.19)--(\x+.40,\y-.19);
}
\foreach \i in {1,2,3}{\node[mass,fill=blue!12] (m\i) at (p\i) {$\i$};}
\foreach \i in {4,5,6,7,8}{\node[mass,fill=violet!14] (m\i) at (p\i) {$\i$};}
\foreach \a/\b in {1/2,2/3,2/4,4/5,4/6,6/7,7/8,1/3,5/7}{\MPCrailLink{m\a}{m\b}}
\draw[-{Latex[length=1.5mm]}] (-.1,-1.6)--(.65,-1.6) node[right,font=\scriptsize] {$q_i,u_i$};
\node[font=\scriptsize,anchor=west] at (-.1,-2.05) {horizontal motion};
\end{tikzpicture}
\endgroup}
\caption{Eight-agent spring--damper network used in
Section~\ref{sec:system}. Masses move horizontally on guides;
each edge represents an axial spring and damper in parallel.
Equilibrium coordinates are
$(0,1),(1,0),(2,1),(2,-1),(3,0),(3,-2),(4,-1),(5,-2)$ m.
Colours illustrate $\Ccal_1=\{1,2,3\}$ and
$\Ccal_2=\{4,5,6,7,8\}$. Ground connections are omitted for clarity.}
\label{fig:ring-graph-clusters}
\end{figure}

To describe cooperative interactions among agents for control purposes, we introduce the notion of \emph{clusters}. In this work, clusters represent groups of agents that coordinate their control actions based on shared model, state, constraint and objective information~\cite{chanfreut2021survey}. More precisely, couplings between agents belonging to the same cluster are accounted for and coordinated in the optimization of the control decisions, whereas couplings between agents in different clusters are handled robustly as bounded, unknown disturbances. In this way, the clustering induces a structured decomposition of the global dynamics into an intra-cluster component and a robust inter-cluster component. An example of a graph split into two clusters is given in Figure~\ref{fig:ring-graph-clusters}.

Formally, let $N_{\mathrm{c}}(k)$ denote the number of clusters at time step $k$, and let each cluster $c \in \mathbb{N}_{[1,N_{\mathrm{c}}(k)]}$ be represented by a set of agent indices
\(
\mathcal{C}_c(k) \subseteq \mathbb{N}_{[1,N]}.
\)
The family of clusters is assumed to form a \emph{partition} or \emph{clustering} \(\Pcal(k) = \{\Ccal_{1}(k),\ldots,\Ccal_{N_c(k)}(k)\}\) of the agent set at time \(k\), in the sense that,
\begin{align*}
\mathcal{C}_{c_1}(k)\cap \mathcal{C}_{c_2}(k)&=\emptyset,\quad \forall c_1\neq c_2, &
\bigcup_{c=1}^{N_{\mathrm{c}}(k)} \mathcal{C}_c(k)&=\mathbb{N}_{[1,N]}.
\end{align*}
The quantity $N_{\mathrm{c}}(k)$ characterizes the degree of decentralization of the control architecture at time step $k$. In particular, $N_{\mathrm{c}}(k)=1$ corresponds to a fully centralized MPC scheme, whereas $N_{\mathrm{c}}(k)=N$ corresponds to a fully decentralized scheme. Intermediate values of $N_{\mathrm{c}}(k)$ represent partially distributed control structures. In this sense, the cluster partition determines both the communication topology and the modeling granularity used for control synthesis. In what follows, when no ambiguity arises, the dependence on the time index $k$ is omitted for notational simplicity. For two agents $i$ and $j$, we write $i \sim j$ if they belong to the same cluster, and $i \nsim j$ otherwise.  

Since the number of partitions of a set of \(N\) agents is given by the \(N\)th Bell number \(B_N\), the number of possible clustering topologies grows rapidly with the network size. Consequently, evaluating every possible partition becomes computationally prohibitive even for networks of moderate size. To preserve tractability, we restrict the analysis to a designer-specified collection of admissible clusters, denoted by \(\Ccal_{\mathrm{adm}}\), and perform the required synthesis computations for these clusters offline. This restriction allows the designer to exclude clusters for which the construction of invariant tubes, tightened constraint sets, or terminal ingredients would be prohibitively expensive because of their dimension. 

The admissible cluster family induces a corresponding set of admissible partitions. Let $\mathbb{P}(\mathcal{N})$ denote the set of all partitions of set $\mathcal{N}$. A partition \(\Pcal\) is called admissible if every cluster appearing in it belongs to \(\Ccal_{\mathrm{adm}}\), that is,
\begin{equation}
    \Pcal_{\mathrm{adm}} := \left\{\Pcal \in\mathbb{P}(\mathcal{N}):
    \forall \Ccal \in \Pcal,\ \Ccal \in \Ccal_{\mathrm{adm}}
    \right\}.
\end{equation}

\subsection{Design Objectives}

The goal of this work is to design a scalable MPC strategy for large-scale coupled multi-agent systems. The strategy should remain computationally tractable, limit state information sharing across agents, and achieve near-centralized performance without requiring full cooperation at all times.

To this end, the proposed approach introduces a \textit{dynamic partition} strategy that splits the agents into clusters, adapting the degree of cooperation online. The scheme encapsulates both centralized and decentralized MPC, as well as any communication structure in-between these two extremes. By limiting cooperation to smaller subsets of the graph network, the computational burden scales with cluster size, rather than the full network. Furthermore, state sharing is restricted to intra-cluster agents, limiting data exposure across the network.

This partitioning is driven by the \emph{cluster mismatch} function, which is computed online and quantifies the performance gap between a given partition and the fully centralized solution. Intuitively, if at a given state, the mismatch function is small, there is a small potential performance difference between coordinating and not coordinating. Similarly, if the mismatch function is large, there is a large potential performance difference. We use this mismatch function to change the partition online.

Hence, this partition switching is based on state-dependent performance impact, rather than network structure alone. The underlying motivation is that centralized and decentralized MPC can achieve similar performance in certain structures, despite vastly different communication requirements. This property is exploited by the proposed approach, to limit information exchange only when it is strictly necessary.

\section{Tube-Based MPC \& The Mismatch Function }\label{sec:tube_mpc}
In this section we formulate the tube-based MPC subproblem that is used by each cluster and introduce the concept of a \emph{cluster mismatch} function.
\subsection{Cluster-Level Tube-Based MPC}
At a given time $k\in \mathbb{Z}_{\geq 0}$, consider a partition
$\mathcal{P}(k)\in P_{\mathrm{adm}}$ and a cluster
$\Ccal \in \Pcal(k)$. The corresponding cluster state and input are
obtained by stacking the variables of all agents belonging to $\Ccal$:
\begin{align}
x_\Ccal(k)
&:= \mathrm{col}_{i \in \Ccal}(x_i(k))
\in \mathbb{R}^{\sum_{i\in \Ccal} n_i}, \\
u_\Ccal(k)
&:= \mathrm{col}_{i \in\Ccal}(u_i(k))
\in \mathbb{R}^{\sum_{i\in \Ccal} m_i}.
\end{align}
Using the local dynamics in \eqref{eq:model}, the dynamics associated
with cluster $\Ccal$ can be written as
\begin{align}
x_\Ccal(k+1)
&=
A_{\Ccal\Ccal}x_\Ccal(k)
+
B_{\Ccal\Ccal}u_\Ccal(k)
+
w_{\Ccal}(k),
\label{eq:clust_dyn}
\end{align}
where the block matrices $A_{\Ccal\Ccal}$ and $B_{\Ccal\Ccal}$ are
given by
\begin{equation}
A_{\Ccal\Ccal}
=
[A_{ij}]_{i,j\in\Ccal},
\qquad
B_{\Ccal\Ccal}
=
[B_{ij}]_{i,j\in\Ccal}.
\end{equation}

To describe the coupling generated by agents outside $\Ccal$, define
the set of external neighbors of the cluster as
\begin{equation}
\Ncal_\Ccal^{\mathrm{ext}}
:=
\left(
\bigcup_{i\in\Ccal}\Ncal_i
\right)
\setminus \Ccal .
\end{equation}
For each $j\in\Ncal_\Ccal^{\mathrm{ext}}$, define the corresponding
stacked coupling matrices
\begin{equation}
A_{\Ccal j}
:=
\mathrm{col}_{i\in\Ccal}(A_{ij}),
\qquad
B_{\Ccal j}
:=
\mathrm{col}_{i\in\Ccal}(B_{ij}),
\label{eq:stacked_external_coupling}
\end{equation}
where $A_{ij}=0$ and $B_{ij}=0$ whenever $j\notin\Ncal_i$.
The inter-cluster coupling term can then be written as
\begin{equation}
w_\Ccal(k)
=
\sum_{j\in\Ncal_\Ccal^{\mathrm{ext}}}
\left(
A_{\Ccal j}x_j(k)
+
B_{\Ccal j}u_j(k)
\right).
\label{eq:clust_dist}
\end{equation}
Hence, the contribution of a given external agent $j$ to all agents
within $\Ccal$ is represented jointly through the same state $x_j$
and input $u_j$.

Since $x_j\in\Xcal_j$ and $u_j\in\Ucal_j$ by
\eqref{eq:constraints_local}, the inter-cluster coupling acts as a
bounded disturbance. An admissible disturbance set is therefore
\begin{equation}
\Wcal_{\Ccal}
:=
\bigoplus_{j\in\Ncal_\Ccal^{\mathrm{ext}}}
\left(
A_{\Ccal j}\Xcal_j
\oplus
B_{\Ccal j}\Ucal_j
\right),
\label{eq:cluster_disturbance_set}
\end{equation}
such that $
w_{\Ccal}(k)\in\Wcal_{\Ccal}
$
for every admissible global state and input. In particular, $\Wcal_{\Ccal}=\{0\}$ when $\Ccal=\Ncal$, corresponding
to centralized operation. The cluster-level state and input constraint sets are consequently given by \begin{equation}
\Xcal_{\Ccal}:=\prod_{i\in\Ccal}\Xcal_i,
\qquad
\Ucal_{\Ccal}:=\prod_{i\in\Ccal}\Ucal_i.
\end{equation}
Furthermore, define
\begin{equation}
Q_{\Ccal}:=\operatorname{blkdiag}_{i\in\Ccal}[Q_i],
\qquad
R_{\Ccal}:=\operatorname{blkdiag}_{i\in\Ccal}[R_i],
\end{equation}
so that the cluster stage cost is
\begin{equation}
\ell_{\Ccal}(\chi_{\Ccal},\nu_{\Ccal})
:=
\chi_{\Ccal}^{\top}Q_{\Ccal}\chi_{\Ccal}
+
\nu_{\Ccal}^{\top}R_{\Ccal}\nu_{\Ccal}
=
\sum_{i\in\Ccal}
\ell_i(\chi_i,\nu_i).
\end{equation}
Following the procedure of classical tube-based MPC~\cite{baldivieso2021tube}, for each admissible cluster, we consider the nominal disturbance-free system
\begin{equation}
z_{\Ccal}(k+1)
=
A_{\Ccal\Ccal}z_{\Ccal}(k)
+
B_{\Ccal\Ccal}v_{\Ccal}(k),
\end{equation}
where $z_{\Ccal}$ and $v_{\Ccal}$ denote the nominal cluster state and input,
respectively. The deviation between the real and nominal cluster states is
defined as the error of the cluster $\Ccal$, given by:
\begin{equation}
e_{\Ccal}(k):=x_{\Ccal}(k)-z_{\Ccal}(k).
\end{equation}
Let $K_{\Ccal}$ denote an auxiliary feedback gain and parameterize the real
cluster input according to
\begin{equation}
u_{\Ccal}(k)
=
v_{\Ccal}(k)
+
K_{\Ccal}e_{\Ccal}(k).
\label{eq:real_input}
\end{equation}
The resulting error dynamics are thus given by
\begin{equation}
e_{\Ccal}(k+1)
=
\left(
A_{\Ccal\Ccal}
+
B_{\Ccal\Ccal}K_{\Ccal}
\right)e_{\Ccal}(k)
+
w_{\Ccal}(k).
\label{eq:error_dyn}
\end{equation}

For each $\Ccal\in\Ccal_{\mathrm{adm}}$, let $\Ecal_{\Ccal}$ be a robust
positively invariant (RPI) set for~\eqref{eq:error_dyn}, i.e., it satisfies the condition
\begin{equation}
\left(
A_{\Ccal\Ccal}
+
B_{\Ccal\Ccal}K_{\Ccal}
\right)\Ecal_{\Ccal}
\oplus
\Wcal_{\Ccal}
\subseteq
\Ecal_{\Ccal}.
\end{equation}
Consequently, for every $w_{\Ccal}(k)\in\Wcal_{\Ccal}$,
\(
e_{\Ccal}(k)\in\Ecal_{\Ccal}
\implies 
e_{\Ccal}(k+1)\in\Ecal_{\Ccal}
\). The real cluster state therefore
remains inside the tube
$
x_{\Ccal}(k)\in z_{\Ccal}(k)\oplus\Ecal_{\Ccal}.
$

To guarantee satisfaction of the original state and input constraints for
every admissible tube error, define the tightened nominal constraint sets
\begin{equation}
\overline{\Xcal}_{\Ccal}
:=
\Xcal_{\Ccal}\ominus\Ecal_{\Ccal},
\qquad
\overline{\Ucal}_{\Ccal}
:=
\Ucal_{\Ccal}\ominus K_{\Ccal}\Ecal_{\Ccal}.
\end{equation}
Indeed, if
\(
z_{\Ccal}(k)\in\overline{\Xcal}_{\Ccal},
e_{\Ccal}(k)\in\Ecal_{\Ccal}\) then \(
x_{\Ccal}(k)=z_{\Ccal}(k)+e_{\Ccal}(k)\in\Xcal_{\Ccal},
\)
and, similarly, if \(
v_{\Ccal}(k)\in\overline{\Ucal}_{\Ccal},
e_{\Ccal}(k)\in\Ecal_{\Ccal} \) then \(
u_{\Ccal}(k)=v_{\Ccal}(k)+K_{\Ccal}e_{\Ccal}(k)\in\Ucal_{\Ccal}.
\)
The auxiliary feedback gain $K_{\Ccal}$ regulates the deviation between
the real and nominal trajectories through the error dynamics
in~\eqref{eq:error_dyn}. 

For every admissible cluster
$\Ccal\in\Ccal_{\mathrm{adm}}$, let
$K_{\mathrm f,\Ccal}$ denote a terminal feedback gain and define
\begin{equation}
    A_{\mathrm f,\Ccal}
    :=
    A_{\Ccal\Ccal}
    +
    B_{\Ccal\Ccal}K_{\mathrm f,\Ccal}.
    \label{eq:terminal_closed_loop}
\end{equation}
The gain $K_{\mathrm f,\Ccal}$ is selected such that
$A_{\mathrm f,\Ccal}$ is Schur stable. Note that the ancillary feedback law $K_\Ccal$ can be distinct from the
feedback law used to stabilize the nominal system inside the terminal
region $K_{\mathrm f,\Ccal}$, but it does not have to be.

Associated with the terminal controller, define the quadratic terminal
cost
\begin{equation}
    V_{\mathrm f,\Ccal}(z_\Ccal)
    :=
    z_\Ccal^\top P_\Ccal z_\Ccal,
    \qquad
    P_\Ccal\succ 0.
    \label{eq:terminal_cost}
\end{equation}
The matrix $P_\Ccal$ is selected such that
\begin{equation}
A_{\mathrm f,\Ccal}^{\top}
P_\Ccal
A_{\mathrm f,\Ccal}
-
P_\Ccal
\preceq
-
\left(
Q_\Ccal
+
K_{\mathrm f,\Ccal}^{\top}
R_\Ccal
K_{\mathrm f,\Ccal}
\right).
\label{eq:terminal_cost_decrease}
\end{equation}
Consequently, under the terminal controller
$v_\Ccal=K_{\mathrm f,\Ccal}z_\Ccal$,
\begin{equation}
V_{\mathrm f,\Ccal}
\left(
A_{\mathrm f,\Ccal}z_\Ccal
\right)
-
V_{\mathrm f,\Ccal}(z_\Ccal)
\leq
-
\ell_\Ccal
\left(
z_\Ccal,
K_{\mathrm f,\Ccal}z_\Ccal
\right).
\label{eq:terminal_lyapunov_decrease}
\end{equation}

Furthermore, let $\Zcal_{\mathrm f,\Ccal}$ denote a compact terminal set
containing the origin such that
\begin{align}
A_{\mathrm f,\Ccal}\Zcal_{\mathrm f,\Ccal}
&\subseteq
\Zcal_{\mathrm f,\Ccal},
\label{eq:terminal_set_invariance}
\\
\Zcal_{\mathrm f,\Ccal}
&\subseteq
\overline{\Xcal}_\Ccal,
\label{eq:terminal_set_state}
\\
K_{\mathrm f,\Ccal}\Zcal_{\mathrm f,\Ccal}
&\subseteq
\overline{\Ucal}_\Ccal.
\label{eq:terminal_set_input}
\end{align}
Hence, once the nominal state enters $\Zcal_{\mathrm f,\Ccal}$,
application of the terminal controller preserves feasibility of the
tightened nominal constraints while decreasing the terminal cost.

In Assumption~\ref{ass:tube_ingredients} we collect the conditions
required for the tube-based MPC construction.

\begin{assumption}
\label{ass:tube_ingredients}
For each $\Ccal\in\Ccal_{\mathrm{adm}}$, the corresponding tube MPC
ingredients satisfy:
\begin{enumerate}
    \item The gain $K_{\Ccal}$ is selected such that $
        A_{\Ccal\Ccal}
        +
        B_{\Ccal\Ccal}K_{\Ccal}$
    is Schur stable.

    \item The set $\Ecal_\Ccal$ is compact, convex, contains the origin,
    and is RPI for the error dynamics.

    \item The tightened state and input constraint sets are nonempty and
    contain the origin.
    \item There exist a terminal feedback gain
$K_{\mathrm f,\Ccal}$, a matrix $P_\Ccal\succ0$, and a compact
terminal set $\Zcal_{\mathrm f,\Ccal}$ satisfying
\(
0\in\operatorname{int}\left(\Zcal_{\mathrm f,\Ccal}\right),
\)
such that
    \eqref{eq:terminal_cost_decrease}--%
    \eqref{eq:terminal_set_input}
hold.
\end{enumerate}
\end{assumption}

We have now the required ingredients to define the online optimization
problem. Given a time $k\in\mathbb{Z}_{\geq 0}$ and prediction horizon
$N_{\mathrm p}$, define the predicted nominal sequences
\begin{equation}
\mathbf z_{\Ccal}(k)
:=
\operatorname{col}_{t=0}^{N_{\mathrm p}}
\left(z_{\Ccal,t|k}\right),
\qquad
\mathbf v_{\Ccal}(k)
:=
\operatorname{col}_{t=0}^{N_{\mathrm p}-1}
\left(v_{\Ccal,t|k}\right).
\end{equation}
The cluster-level tube-based MPC problem at time $k$ is
\begin{subequations}
\label{eq:nom_opt_C}
\begin{align}
J_{\Ccal}^{\star}(x_{\Ccal}(k))
&:=
\min_{\substack{
\mathbf z_{\Ccal}(k),\\
\mathbf v_{\Ccal}(k)}}
\left[
\sum_{t=0}^{N_{\mathrm p}-1}
\ell_{\Ccal}
\left(
z_{\Ccal,t|k},
v_{\Ccal,t|k}
\right)
+
V_{\mathrm f,\Ccal}
\left(
z_{\Ccal,N_{\mathrm p}|k}
\right)
\right]
\label{eq:nom_opt_C_cost}
\\
\mathrm{s.t.} \quad&
x_{\Ccal}(k)-z_{\Ccal,0|k}
\in
\Ecal_{\Ccal},
\label{eq:initial_tube}
\\
&
z_{\Ccal,t+1|k}
=
A_{\Ccal\Ccal}z_{\Ccal,t|k}
+
B_{\Ccal\Ccal}v_{\Ccal,t|k}, \quad
t\in\mathbb{N}_{[0,N_{\mathrm p}-1]},
\\
&
z_{\Ccal,t|k}
\in
\overline{\Xcal}_{\Ccal}, \qquad
t\in\mathbb{N}_{[0,N_{\mathrm p}-1]},
\\
&
v_{\Ccal,t|k}
\in
\overline{\Ucal}_{\Ccal}, \qquad
t\in\mathbb{N}_{[0,N_{\mathrm p}-1]},
\\
&
z_{\Ccal,N_{\mathrm p}|k}
\in
\Zcal_{\mathrm f,\Ccal}.
\label{eq:terminal}
\end{align}
\end{subequations}
Let \(
\left(
\mathbf z_{\Ccal}^{\star}(k),
\mathbf v_{\Ccal}^{\star}(k)
\right)
\)
denote an optimal solution of~\eqref{eq:nom_opt_C}. From \eqref{eq:real_input}, the input
implemented by the agents in cluster $\Ccal$ is
\begin{equation}
u_{\Ccal}(k)
=
v_{\Ccal,0|k}^{\star}
+
K_{\Ccal}
\left(
x_{\Ccal}(k)-z_{\Ccal,0|k}^{\star}
\right).
\label{eq:u_implement}
\end{equation}

The initial tube constraint
\eqref{eq:initial_tube} allows the nominal initial state to differ
from the measured state. Robust positive invariance of $\Ecal_{\Ccal}$
then ensures that every subsequent error term remains within the tube
\(
x_{\Ccal}(k+1)-z_{\Ccal,1|k}^{\star}
\in
\Ecal_{\Ccal}
\), for every $w_{\Ccal}(k)\in\Wcal_{\Ccal}$.

All tube, tightening, and terminal ingredients
\begin{equation}
  \left(
K_{\Ccal},
\Ecal_{\Ccal},
\overline{\Xcal}_{\Ccal},
\overline{\Ucal}_{\Ccal},
K_{\mathrm f,\Ccal},
P_\Ccal,
\Zcal_{\mathrm f,\Ccal}
\right)  
\end{equation}
are computed offline for every
$\Ccal\in\Ccal_{\mathrm{adm}}$. During online operation, the
controller therefore only solves the nominal quadratic program
in~\eqref{eq:nom_opt_C}. Agents belonging to the same cluster
coordinate their internal couplings through the matrices
$A_{\Ccal\Ccal}$ and $B_{\Ccal\Ccal}$, whereas couplings generated
by agents outside the cluster are handled robustly through the
disturbance set $\Wcal_{\Ccal}$ and invariant tube $\Ecal_{\Ccal}$.

\subsection{The Upper Bound}
We next quantify the predicted performance impact associated with a
given cluster. To this end, we construct two deliberately different
bounds. First, the tube-MPC solution is used to obtain a robust upper
bound on the real predicted cluster cost. Second, we construct an
optimistic lower bound by explicitly finding those disturbances that minimize the local predicted costs. The difference between these quantities
defines the cluster mismatch function.

To achieve this, we begin by distinguishing between the nominal predictions computed by the cluster-level MPC problems and the corresponding real predicted trajectories under the ancillary feedback law. For simplicity, we first consider the case in which we operate in a decentralized manner, that is $\Ccal_i = \{i\}$, $N_c = N$. For each agent $i$, let
\begin{equation}
  \mathbf{z}^\star_{i}(k) =  \begin{bmatrix}   z^\star_{i,0|k} \\ z^\star_{i,1|k} \\ \vdots \\ z^\star_{i,N_\mathrm{p}|k}
    \end{bmatrix}, \quad \mathbf{v}^\star_{i}(k) =  \begin{bmatrix}   v^\star_{i,0|k} \\ v^\star_{i,1|k} \\ \vdots \\ v^\star_{i,N_\mathrm{p}-1|k}
    \end{bmatrix},  \label{eq:nominal_sequences}
\end{equation}
denote the optimal nominal state and input sequences obtained from problem~\eqref{eq:nom_opt_C}. First define the predicted error sequence through $\mathbf{e}_i = [e_{i,t|k}]_{t=0}^{N_p} $, $e_{i,t|k} = x_{i,t|k}^\star - z_{i,t|k}^\star$. The corresponding \emph{real} predicted state and input sequences are given by
\begin{align}
    \mathbf{x}^\star_{i}(k) &=  \begin{bmatrix}   x^\star_{i,0|k} \\ x^\star_{i,1|k} \\ \vdots \\ x^\star_{i,N_\mathrm{p}|k}
    \end{bmatrix} = \begin{bmatrix}   z^\star_{i,0|k} + e_{i, 0|k}\\ z^\star_{i,1|k} + e_{i, 1|k} \\ \vdots \\ z^\star_{i,N_\mathrm{p}|k} + e_{i, N_\mathrm{p}|k}
    \end{bmatrix}, \\
    \mathbf{u}^\star_{i}(k) &=  \begin{bmatrix}   u^\star_{i,0|k} \\ u^\star_{i,1|k} \\ \vdots \\ u^\star_{i,N_\mathrm{p}-1|k}
    \end{bmatrix}  = \begin{bmatrix}   v^\star_{i,0|k}  + K_i e_{i, 0|k} \\ v^\star_{i,1|k} + K_i e_{i, 1|k} \\ \vdots \\ v^\star_{i,N_\mathrm{p}-1|k} + K_i e_{i, N_\mathrm{p}-1|k}
    \end{bmatrix},
    \label{eq:real_sequences}
\end{align}
where $K_i$ is abuse of notation for $K_{\{i\}}$. Accordingly, define the $N_{\mathrm p}$-step real predicted stage cost
of agent $i$ as
\begin{equation}
J_i^{\mathrm r}\left(
\mathbf{x}_i^\star(k),
\mathbf{u}_i^\star(k)
\right)
:=
\sum_{t=0}^{N_{\mathrm p}-1}
\ell_i
\left(
x^\star_{i,t|k},
u^\star_{i,t|k}
\right),
\label{eq:real_cost}
\end{equation}
and the corresponding nominal predicted stage cost as
\begin{equation}
J_i^{\mathrm n}\left(
\mathbf{z}_i^\star(k),
\mathbf{v}_i^\star(k)
\right)
:=
\sum_{t=0}^{N_{\mathrm p}-1}
\ell_i
\left(
z^\star_{i,t|k},
v^\star_{i,t|k}
\right).
\label{eq:nominal_cost}
\end{equation}
The nominal quantity in~\eqref{eq:nominal_cost} contains only the
$N_{\mathrm p}$-step stage cost and therefore differs from the objective
of problem~\eqref{eq:nom_opt_C}, which additionally contains the
terminal cost. For later use, define the corresponding augmented nominal
cost as
\begin{equation}
\Phi_\Ccal
\left(
\mathbf z_\Ccal,
\mathbf v_\Ccal
\right)
:=
J_\Ccal^{\mathrm n}
\left(
\mathbf z_\Ccal,
\mathbf v_\Ccal
\right)
+
V_{\mathrm f,\Ccal}
\left(
z_{\Ccal,N_{\mathrm p}}
\right).
\label{eq:augmented_nominal_cost}
\end{equation}
For the optimal solution of problem~\eqref{eq:nom_opt_C}, the resulting sequences $\left(
\mathbf z_\Ccal^\star(k),
\mathbf v_\Ccal^\star(k)
\right)$ define the optimal cost
\begin{equation}
J_\Ccal^\star(x_\Ccal(k))
=
\Phi_\Ccal
\left(
\mathbf z_\Ccal^\star(k),
\mathbf v_\Ccal^\star(k)
\right).
\label{eq:optimal_augmented_cost}
\end{equation}
Moreover, since
$V_{\mathrm f,\Ccal}(z_\Ccal)\geq0$,
\(
J_\Ccal^{\mathrm n}
\left(
\mathbf z_\Ccal,
\mathbf v_\Ccal
\right)
\leq
\Phi_\Ccal
\left(
\mathbf z_\Ccal,
\mathbf v_\Ccal
\right).
\)

We next relate the real and nominal predicted stage costs. To simplify
notation, we temporarily omit the time index $k$ and superscript
$\star$. Using \(
x_{i,t}=z_{i,t}+e_{i,t}, u_{i,t}=v_{i,t}+K_i e_{i,t},
\)
we obtain
\begin{align}
J_i^{\mathrm r}\left(
\mathbf{x}_i^\star(k),
\mathbf{u}_i^\star(k)
\right)
&=
J^{\mathrm{n}}_i\left(
\mathbf{z}_i^\star(k),
\mathbf{v}_i^\star(k)
\right)
\notag\\
&\quad+
\sum_{t=0}^{N_{\mathrm p}-1}
e_{i,t}^{\top}
\left(
Q_i+K_i^\top R_iK_i
\right)e_{i,t}
\notag\\
&\quad+
2\sum_{t=0}^{N_{\mathrm p}-1}
z_{i,t}^{\top}Q_i e_{i,t}
+
2\sum_{t=0}^{N_{\mathrm p}-1}
v_{i,t}^{\top}R_iK_i e_{i,t}.
\label{eq:cost_expansion}
\end{align}
Using Young's inequality, which states that for $G\in \mathbb{R}^{n\times n}$, $a,b\in\mathbb{R}^n$,
\begin{equation}
2a^\top G b
\leq
\varepsilon a^\top G a
+
\frac{1}{\varepsilon}b^\top G b,
\qquad
\varepsilon>0,
\label{eq:young}
\end{equation}
we obtain that the cross terms satisfy
\begin{align}
2z_{i,t}^\top Q_i e_{i,t}
&\leq
\varepsilon z_{i,t}^\top Q_i z_{i,t}
+
\frac{1}{\varepsilon}e_{i,t}^\top Q_i e_{i,t},
\\
2v_{i,t}^\top R_iK_i e_{i,t}
&\leq
\varepsilon v_{i,t}^\top R_i v_{i,t}
+
\frac{1}{\varepsilon}
e_{i,t}^\top K_i^\top R_iK_i e_{i,t}.
\end{align}
Consequently,
\begin{equation}
J_i^{\mathrm r}(\mathbf{x}_i,\mathbf{u}_i)
\leq
(1+\varepsilon)
J_i^{\mathrm n}(\mathbf{z}_i,\mathbf{v}_i)
+
\left(1+\frac{1}{\varepsilon}\right)
\mathbf e_i^\top M_i\mathbf e_i,
\label{eq:local_bound}
\end{equation}
where now
\(
M_i
:=
\operatorname{blkdiag}_{t=0}^{N_{\mathrm p}-1}
\left(
Q_i+K_i^\top R_iK_i
\right),
\label{eq:Mi}
\)
and
\(
\mathbf e_i
:=
\operatorname{col}_{t=0}^{N_{\mathrm p}-1}(e_{i,t}).
\)

The same argument extends directly to any admissible cluster
$\Ccal\in\Ccal_{\mathrm{adm}}$. Define
\begin{equation}
J_{\Ccal}^{\mathrm r}
(\mathbf{x}_{\Ccal},\mathbf{u}_{\Ccal})
:=
\sum_{t=0}^{N_{\mathrm p}-1}
\ell_{\Ccal}(x_{\Ccal,t},u_{\Ccal,t})
\end{equation}
and
\begin{equation}
J_{\Ccal}^{\mathrm n}
(\mathbf{z}_{\Ccal},\mathbf{v}_{\Ccal})
:=
\sum_{t=0}^{N_{\mathrm p}-1}
\ell_{\Ccal}(z_{\Ccal,t},v_{\Ccal,t}).
\end{equation}
Then
\begin{equation}
J_{\Ccal}^{\mathrm r}
(\mathbf{x}_{\Ccal}^{\star},\mathbf{u}_{\Ccal}^{\star})
\leq
(1+\varepsilon)
J_{\Ccal}^{\mathrm n}
(\mathbf{z}_{\Ccal}^{\star},\mathbf{v}_{\Ccal}^{\star})
+
\left(1+\frac{1}{\varepsilon}\right)
\mathbf e_{\Ccal}^\top
M_{\Ccal}
\mathbf e_{\Ccal},
\label{eq:cluster_bound_optimal}
\end{equation}
where
\(
M_{\Ccal}
:=
\operatorname{blkdiag}_{t=0}^{N_{\mathrm p}-1}
\left(
Q_{\Ccal}+K_{\Ccal}^{\top}R_{\Ccal}K_{\Ccal}
\right).
\)

Let
$(\tilde{\mathbf z}_{\Ccal}(k),\tilde{\mathbf v}_{\Ccal}(k))$
be any feasible candidate for problem~\eqref{eq:nom_opt_C}.
Since the MPC problem minimizes the augmented nominal cost
$\Phi_\Ccal$ defined in~\eqref{eq:augmented_nominal_cost}, rather than
the stage cost $J_\Ccal^{\mathrm n}$ alone, optimality gives
\begin{equation}
\Phi_\Ccal
\left(
\mathbf z_\Ccal^\star(k),
\mathbf v_\Ccal^\star(k)
\right)
=
J_\Ccal^\star(x_\Ccal(k))
\leq
\Phi_\Ccal
\left(
\tilde{\mathbf z}_\Ccal(k),
\tilde{\mathbf v}_\Ccal(k)
\right).
\label{eq:optimal_augmented_candidate}
\end{equation}
Furthermore, since the terminal cost is nonnegative,
\begin{equation}
J_\Ccal^{\mathrm n}
\left(
\mathbf z_\Ccal^\star(k),
\mathbf v_\Ccal^\star(k)
\right)
\leq
\Phi_\Ccal
\left(
\mathbf z_\Ccal^\star(k),
\mathbf v_\Ccal^\star(k)
\right).
\end{equation}
Combining these relations with~\eqref{eq:cluster_bound_optimal}
therefore yields
\begin{align}
J_\Ccal^{\mathrm r}
\left(
\mathbf x_\Ccal^\star(k),
\mathbf u_\Ccal^\star(k)
\right)
&\leq
(1+\varepsilon)
\Phi_\Ccal
\left(
\tilde{\mathbf z}_\Ccal(k),
\tilde{\mathbf v}_\Ccal(k)
\right)
\notag\\
&\quad+
\left(
1+\frac{1}{\varepsilon}
\right)
\mathbf e_\Ccal^\top
M_\Ccal
\mathbf e_\Ccal,
\qquad
\varepsilon>0.
\label{eq:cluster_bound_candidate}
\end{align}

This expression separates the contribution associated with the
candidate nominal trajectory from that associated with the tube error.
Since for $t\in\mathbb N_{[0,N_{\mathrm p}-1]}$,
\(
e_{\Ccal,t|k}\in\Ecal_\Ccal,
\)
define the worst-case tube contribution
\begin{equation}
\Gamma_\Ccal
:=
\max_{\substack{
e_{\Ccal,t}\in\Ecal_\Ccal\\
t=0,\ldots,N_{\mathrm p}-1}}
\sum_{t=0}^{N_{\mathrm p}-1}
e_{\Ccal,t}^{\top}
\left(
Q_\Ccal
+
K_\Ccal^\top R_\Ccal K_\Ccal
\right)
e_{\Ccal,t}.
\label{eq:GammaC}
\end{equation}
Compactness of $\Ecal_\Ccal$ guarantees that
$\Gamma_\Ccal<\infty$. Moreover, $\Gamma_\Ccal$ depends only on the
offline-computed tube ingredients of cluster $\Ccal$ and can therefore
be evaluated offline.

It follows from~\eqref{eq:cluster_bound_candidate} that, for every
$\varepsilon>0$,
\begin{align}
J_\Ccal^{\mathrm r}
\left(
\mathbf x_\Ccal^\star(k),
\mathbf u_\Ccal^\star(k)
\right)
&\leq
(1+\varepsilon)
\Phi_\Ccal
\left(
\tilde{\mathbf z}_\Ccal(k),
\tilde{\mathbf v}_\Ccal(k)
\right)
\notag\\
&\quad+
\left(
1+\frac{1}{\varepsilon}
\right)
\Gamma_\Ccal.
\label{eq:real_cost_bound_epsilon}
\end{align}
Accordingly, for a given $\varepsilon>0$, define
\begin{align}
\bar J_{\Ccal,\varepsilon}
\left(
\tilde{\mathbf z}_\Ccal(k),
\tilde{\mathbf v}_\Ccal(k)
\right)
&:=
(1+\varepsilon)
\Phi_\Ccal
\left(
\tilde{\mathbf z}_\Ccal(k),
\tilde{\mathbf v}_\Ccal(k)
\right)
\notag\\
&\quad+
\left(
1+\frac{1}{\varepsilon}
\right)
\Gamma_\Ccal.
\label{eq:upper_bound_epsilon}
\end{align}
Hence, $\forall\varepsilon>0$ we have
\begin{equation}
J_\Ccal^{\mathrm r}
\left(
\mathbf x_\Ccal^\star(k),
\mathbf u_\Ccal^\star(k)
\right)
\leq
\bar J_{\Ccal,\varepsilon}
\left(
\tilde{\mathbf z}_\Ccal(k),
\tilde{\mathbf v}_\Ccal(k)
\right),
 \label{eq:algorithm_upper_bound}
\end{equation}
The parameter $\varepsilon$ originates only from Young's inequality and
can be selected so as to tighten the upper bound. For compactness, let
\begin{equation}
p_\Ccal
:=
\Phi_\Ccal
\left(
\tilde{\mathbf z}_\Ccal(k),
\tilde{\mathbf v}_\Ccal(k)
\right),
\qquad
q_\Ccal
:=
\Gamma_\Ccal.
\end{equation}
Then~\eqref{eq:upper_bound_epsilon} can be written as
\begin{equation}
\bar J_{\Ccal,\varepsilon}
=
p_\Ccal+q_\Ccal
+
\varepsilon p_\Ccal
+
\frac{q_\Ccal}{\varepsilon}.
\label{eq:eps_form}
\end{equation}
For $p_\Ccal>0$ and $q_\Ccal>0$, the arithmetic--geometric mean
inequality gives
\begin{equation}
\varepsilon p_\Ccal
+
\frac{q_\Ccal}{\varepsilon}
\geq
2\sqrt{p_\Ccal q_\Ccal},
\end{equation}
with equality for
\begin{equation}
\varepsilon_\Ccal^\star
=
\sqrt{
\frac{q_\Ccal}{p_\Ccal}
}
=
\sqrt{
\frac{
\Gamma_\Ccal
}{
\Phi_\Ccal
\left(
\tilde{\mathbf z}_\Ccal(k),
\tilde{\mathbf v}_\Ccal(k)
\right)
}
}.
\label{eq:eps_star}
\end{equation}
We therefore define the optimized upper bound as
\begin{align}
\bar J_\Ccal
\left(
\tilde{\mathbf z}_\Ccal(k),
\tilde{\mathbf v}_\Ccal(k)
\right)
&:=
\inf_{\varepsilon>0}
\bar J_{\Ccal,\varepsilon}
\left(
\tilde{\mathbf z}_\Ccal(k),
\tilde{\mathbf v}_\Ccal(k)
\right)
\notag\\
&=
p_\Ccal+q_\Ccal
+
2\sqrt{p_\Ccal q_\Ccal}
\notag\\
&=
\left(
\sqrt{
\Phi_\Ccal
\left(
\tilde{\mathbf z}_\Ccal(k),
\tilde{\mathbf v}_\Ccal(k)
\right)
}
+
\sqrt{\Gamma_\Ccal}
\right)^2.
\label{eq:upper_bound_problem}
\end{align}
For $p_\Ccal>0$ and $q_\Ccal>0$, the infimum is attained at
$\varepsilon_\Ccal^\star$ in~\eqref{eq:eps_star};~\eqref{eq:upper_bound_problem} extends continuously to
$p_\Ccal=0$ or $q_\Ccal=0$.

Since~\eqref{eq:real_cost_bound_epsilon} holds for every
$\varepsilon>0$, the optimized quantity remains an upper bound on the
real predicted cluster cost:
\begin{equation}
J_\Ccal^{\mathrm r}
\left(
\mathbf x_\Ccal^\star(k),
\mathbf u_\Ccal^\star(k)
\right)
\leq
\bar J_\Ccal
\left(
\tilde{\mathbf z}_\Ccal(k),
\tilde{\mathbf v}_\Ccal(k)
\right).
\label{eq:upper_bound}
\end{equation}

\begin{remark}
The upper bound~\eqref{eq:upper_bound_problem} separates the
state-dependent candidate cost from the cluster-dependent tube
contribution. In particular, $\Gamma_\Ccal$ captures the effect of the
disturbance set $\Wcal_\Ccal$ through the invariant tube
$\Ecal_\Ccal$ and is computed offline. Consequently, online evaluation
of the upper bound only requires the augmented nominal candidate cost
\(
\Phi_\Ccal
\left(
\tilde{\mathbf z}_\Ccal(k),
\tilde{\mathbf v}_\Ccal(k)
\right).
\)
Moreover, the closed-form minimization over $\varepsilon$ removes the
need to select the Young parameter as an additional tuning variable.
\end{remark}

\begin{remark}
    Because $M_\Ccal$ is block diagonal across the prediction horizon, the maximization over $\mathbf{e}_\Ccal$ in~\eqref{eq:GammaC} can be decomposed over each block. Second, since the invariant set $\Ecal_\Ccal$ and matrix $M_\Ccal$ are fixed for a given cluster, the error contribution can be precomputed offline, leaving only the nominal term $J_\Ccal(\tilde{\mathbf{z}}_\Ccal(k),\tilde{\mathbf{v}}_\Ccal(k))$ to be evaluated online. The error contribution can instead be stored as:
\(
        \Gamma_\Ccal = \max_{e_\Ccal}\;\mathbf{e}_\Ccal^\top M_\Ccal \mathbf{e}_\Ccal.
\)
\end{remark}

\subsection{The Lower Bound}

Now that the upper bound has been defined, we construct a
partition-independent lower bound on the achievable real predicted
performance. The purpose of this lower bound is deliberately optimistic:
the coupling generated by the neighboring agents is allowed to take,
at each prediction instant, whichever admissible value minimizes the
local cost.

Using the model dynamics~\eqref{eq:model} and disturbance constraint~\eqref{eq:wi_bounds} we define the predicted disturbance sequence of agent $i$ as
\begin{equation}
\mathbf w_i(k)
:=
\operatorname{col}_{t=0}^{N_{\mathrm p}-1}
\left(
w_{i,t|k}
\right),
\qquad
w_{i,t|k}\in\Wcal_i.
\end{equation}

The locally best-case predicted cost is obtained by allowing both the
local control input and the neighboring coupling sequence to take their
most favorable admissible values:
\begin{subequations}
\label{prob:local-best-case}
\begin{align}
J_i^{\min}(x_i(k))
&:=
\min_{\substack{
\mathbf u_i(k),\\
\mathbf w_i(k)}}
\sum_{t=0}^{N_{\mathrm p}-1}
\left(
\xi_{i,t|k}^{\top}Q_i\xi_{i,t|k}
+
u_{i,t|k}^{\top}R_i u_{i,t|k}
\right)
\label{prob:local-best-case-cost}
\\
\mathrm{s.t.}\quad
&
\xi_{i,0|k}
=
x_i(k),
\label{prob:local-best-case-initial}
\\
&
\xi_{i,t+1|k}
=
A_{ii}\xi_{i,t|k}
+
B_{ii}u_{i,t|k}
+
w_{i,t|k},
\notag\\[-1mm]
&\hspace{38mm}
t\in\mathbb N_{[0,N_{\mathrm p}-1]},
\label{prob:local-best-case-dynamics}
\\
&
\xi_{i,t|k}\in\Xcal_i,
\qquad
t\in\mathbb N_{[0,N_{\mathrm p}-1]},
\label{prob:local-best-case-state}
\\
&
u_{i,t|k}\in\Ucal_i,
\qquad
t\in\mathbb N_{[0,N_{\mathrm p}-1]},
\label{prob:local-best-case-input}
\\
&
w_{i,t|k}\in\Wcal_i,
\qquad
t\in\mathbb N_{[0,N_{\mathrm p}-1]}.
\label{prob:local-best-case-disturbance}
\end{align}
\end{subequations}

Importantly, problem~\eqref{prob:local-best-case} does not require the
optimized disturbance sequence $\mathbf w_i(k)$ to be generated by a
dynamically consistent set of neighboring trajectories. Instead, at
every prediction instant, the neighbors are effectively allowed to
coordinate so as to generate whichever value in $\Wcal_i$ is most
favorable to agent $i$. Therefore, $J_i^{\min}$ represents an
optimistic best-case benchmark rather than the performance of a
realizable distributed controller.

Since $Q_i\succ0$ and $R_i\succ0$, the objective in
\eqref{prob:local-best-case} is convex. Moreover, the dynamics are
affine and the sets $\Xcal_i$, $\Ucal_i$, and $\Wcal_i$ are convex
polytopes. Hence, \eqref{prob:local-best-case} is a convex quadratic
program. Compactness of these sets guarantees that, whenever the
problem is feasible, its minimum is attained.

\begin{lemma}
\label{lem:local_lower_bound}
For every agent $i\in \mathcal{N}$, the quantity $J_i^{\min}(x_i(k))$ defined
in~\eqref{prob:local-best-case} is a lower bound on the
$N_{\mathrm p}$-step stage cost of every feasible real predicted
trajectory of agent $i$ initialized at $x_i(k)$:
\begin{equation}
J_i^{\min}(x_i(k))
\leq
J_i^{\mathrm r}(\mathbf x_i,\mathbf u_i).
\label{eq:lower_real}
\end{equation}
\end{lemma}

\begin{proof}
Consider any feasible real predicted trajectory
$(\mathbf x_i,\mathbf u_i)$ initialized at $x_i(k)$. Its dynamics
satisfy
\begin{equation}
x_{i,t+1|k}
=
A_{ii}x_{i,t|k}
+
B_{ii}u_{i,t|k}
+
w_{i,t|k}^{\mathrm r},
\end{equation}
where the realized neighboring coupling is
\begin{equation}
w_{i,t|k}^{\mathrm r}
:=
\sum_{j\in\Ncal_i}
\left(
A_{ij}x_{j,t|k}
+
B_{ij}u_{j,t|k}
\right).
\end{equation}
Since the neighboring state and input trajectories satisfy
$x_{j,t|k}\in\Xcal_j$ and $u_{j,t|k}\in\Ucal_j$, it follows directly
from the definition of $\Wcal_i$ that
\(
w_{i,t|k}^{\mathrm r}\in\Wcal_i.
\)
Consequently, the choices
\[
\xi_{i,t|k}=x_{i,t|k},
\qquad
u_{i,t|k}=u_{i,t|k},
\qquad
w_{i,t|k}=w_{i,t|k}^{\mathrm r}
\]
constitute a feasible solution of
problem~\eqref{prob:local-best-case}. Since
\eqref{prob:local-best-case} minimizes over all such admissible local
input and disturbance sequences, we have obtain \eqref{eq:lower_real}.
\end{proof}

The agent-wise lower bounds can be aggregated to obtain a
partition-independent lower bound for any cluster. Define
\begin{equation}
\underline J_\Ccal(x_\Ccal(k))
:=
\sum_{i\in\Ccal}
J_i^{\min}(x_i(k)).
\label{eq:cluster_lower_bound}
\end{equation}
By Lemma~\ref{lem:local_lower_bound}, for every feasible real predicted
cluster trajectory,
\begin{equation}
\underline J_\Ccal(x_\Ccal(k))
\leq
J_\Ccal^{\mathrm r}
(\mathbf x_\Ccal,\mathbf u_\Ccal).
\label{eq:cluster_lower_real}
\end{equation}

\subsection{The Cluster Mismatch Function}
We are now able to define the cluster mismatch function. Combining~\eqref{eq:cluster_lower_real} with the robust upper bound
in~\eqref{eq:upper_bound} gives
\begin{equation}
\underline J_\Ccal(x_\Ccal(k))
\leq
J_\Ccal^{\mathrm r}
(\mathbf x_\Ccal^\star(k),\mathbf u_\Ccal^\star(k))
\leq
\bar J_\Ccal
(\tilde{\mathbf z}_\Ccal(k),\tilde{\mathbf v}_\Ccal(k)).
\label{eq:sandwich_bound}
\end{equation}
We therefore define the cluster mismatch as
\begin{equation}
\delta_\Ccal(x_\Ccal(k);\tilde{\mathbf z}_\Ccal(k),\tilde{\mathbf v}_\Ccal(k) )
:=
\bar J_\Ccal
(\tilde{\mathbf z}_\Ccal(k),\tilde{\mathbf v}_\Ccal(k))
-
\underline J_\Ccal(x_\Ccal(k)).
\label{eq:deltaC}
\end{equation}
The quantity $\delta_\Ccal$ measures the difference between a robust
upper bound on the predicted real performance associated with cluster
$\Ccal$ and an optimistic best-case benchmark in which the couplings
generated by all neighboring agents are allowed to act in the most
favorable admissible manner. The lower-bound construction is
partition-independent and does not depend on the invariant tube.

\section{Dynamic Clustering Algorithm}
\label{sec:guarantees}

The mismatch function derived in Section~\ref{sec:tube_mpc} is now used
to select the clustering topology online. We first introduce the
switching criterion and its candidate construction, and subsequently
establish performance, recursive feasibility, and practical stability.

\subsection{Switching Criterion}

To balance control performance and communication effort, let
$f_{\mathrm{com}}(\Ccal)$ denote the communication penalty associated
with cluster $\Ccal\in\Ccal_{\mathrm{adm}}$, and define
\begin{equation}
V_\Ccal(x_\Ccal(k))
:=
\delta_\Ccal(x_\Ccal(k);\tilde{\mathbf z}_\Ccal(k),\tilde{\mathbf v}_\Ccal(k))
+
f_{\mathrm{com}}(\Ccal).
\label{eq:cluster-value}
\end{equation}
A possible choice is
\begin{equation}
f_{\mathrm{com}}(\Ccal)=\lambda\phi(\Ccal),
\label{eq:communication-penalty}
\end{equation}
where $\lambda\geq0$ weights communication and $\phi(\Ccal)$ is an
application-dependent measure such as the number of active intra-cluster
communication links.

At time $k\in\mathbb{Z}_{\geq 0}$, the partition is selected according to
\begin{equation}
\Pcal(k)
\in
\arg\min_{\Pcal\in\Pcal_{\mathrm{adm}}}
\sum_{\Ccal\in\Pcal}V_\Ccal(x_\Ccal(k)).
\label{eq:partition-selection}
\end{equation}
Evaluation of $V_\Ccal$ requires a feasible candidate
$(\tilde{\mathbf z}_\Ccal(k),\tilde{\mathbf v}_\Ccal(k))$ for
problem~\eqref{eq:nom_opt_C}. If no such candidate is available, we set
$V_\Ccal(x_\Ccal(k))=\infty$.

For clusters belonging to the previously active partition, a candidate
is available directly from the preceding optimal MPC solution.
Specifically, for $k\geq1$ and $\Ccal\in\Pcal(k-1)$, define the
\emph{shifted candidate}
\begin{align}
\hat z_{\Ccal,t|k}
&:=
z^\star_{\Ccal,t+1|k-1},
&&t=0,\ldots,N_{\mathrm p}-1,
\label{eq:shifted_candidate}
\\
\hat v_{\Ccal,t|k}
&:=
v^\star_{\Ccal,t+1|k-1},
&&t=0,\ldots,N_{\mathrm p}-2,
\nonumber\\
\hat v_{\Ccal,N_{\mathrm p}-1|k}
&:=
K_{\mathrm f,\Ccal}z^\star_{\Ccal,N_{\mathrm p}|k-1},
&&
\hat z_{\Ccal,N_{\mathrm p}|k}
:=
A_{\mathrm f,\Ccal}z^\star_{\Ccal,N_{\mathrm p}|k-1}\nonumber.
\end{align}
After applying~\eqref{eq:u_implement} at time $k-1$, robust positive
invariance of $\Ecal_\Ccal$ gives
\(
x_\Ccal(k)-z^\star_{\Ccal,1|k-1}\in\Ecal_\Ccal.
\)
The remaining constraints are inherited from the preceding optimal
solution and the terminal conditions
\eqref{eq:terminal_set_invariance}--\eqref{eq:terminal_set_input}.
Hence~\eqref{eq:shifted_candidate} is feasible for
problem~\eqref{eq:nom_opt_C} at $x_\Ccal(k)$. Moreover, the terminal
decrease condition~\eqref{eq:terminal_lyapunov_decrease} yields
\begin{equation}
\Phi_\Ccal
\left(
\hat{\mathbf z}_\Ccal(k),
\hat{\mathbf v}_\Ccal(k)
\right)
\leq
J_\Ccal^\star(x_\Ccal(k-1))
-
\ell_\Ccal
\left(
z^\star_{\Ccal,0|k-1},
v^\star_{\Ccal,0|k-1}
\right).
\label{eq:shifted_candidate_decrease}
\end{equation}
Thus the previously active partition always provides a feasible
candidate at the next sampling instant. For clusters outside
$\Pcal(k-1)$, any available feasible candidate may be used. The resulting procedure is summarized in
Algorithm~\ref{alg:dyn}.

\begin{algorithm}[t]
\caption{Dynamic Clustering}
\label{alg:dyn}
\begin{algorithmic}[1]
\STATE \textbf{Input:}
$\Ccal_{\mathrm{adm}}$, $\Pcal_{\mathrm{adm}}$ and offline tube data
\STATE \textbf{Initialize:}
feasible $\Pcal_{\mathrm{ini}}\in\Pcal_{\mathrm{adm}}$ and corresponding
candidate solutions
$P(0) = P_\mathrm{ini}$ 
\FOR{$k=0,1,2,\ldots$}
    \FORALL{$\Ccal\in\Pcal(k)$}
        \STATE Solve~\eqref{eq:nom_opt_C} and apply~\eqref{eq:u_implement}
    \ENDFOR
    \FORALL{$\Ccal\in\Ccal_{\mathrm{adm}}$}
        \STATE Measure $x_\Ccal(k)$ and obtain candidate
        $(\hat{\mathbf z}_\Ccal,\hat{\mathbf v}_\Ccal)$;
        use~\eqref{eq:shifted_candidate} if
         $\Ccal\in\Pcal(k)$
        \IF{the candidate is feasible for~\eqref{eq:nom_opt_C}}
            \STATE
            $V_\Ccal\leftarrow
            \bar J_\Ccal(\hat{\mathbf z}_\Ccal,\hat{\mathbf v}_\Ccal)
            -\underline J_\Ccal(x_\Ccal(k))
            +f_{\mathrm{com}}(\Ccal)$
        \ELSE
            \STATE $V_\Ccal\leftarrow\infty$
        \ENDIF
    \ENDFOR
    \STATE Select $\Pcal(k+1)$ according to~\eqref{eq:partition-selection}
\ENDFOR
\end{algorithmic}
\end{algorithm}

\subsection{Theoretical Guarantees}

We first recall the standard fixed-partition properties of the
cluster-level tube MPC problem. For the remainder of this section, we write $\delta_\Ccal(x_\Ccal(x))$ for $\delta_\Ccal(x_\Ccal(x);\hat{\mathbf z}_\Ccal(k),\hat{\mathbf v}_\Ccal(k))$ with $\hat{\mathbf z}_\Ccal(k),\hat{\mathbf v}_\Ccal(k)$ as obtained by Algorithm~\ref{alg:dyn} at time $k\in\mathbb{Z}_{\geq 0}$. 

\begin{definition}
The closed-loop system is \emph{practically asymptotically stable} if
there exists $\eta\geq0$ such that every feasible trajectory satisfies
\begin{equation}
\limsup_{k\to\infty}\|x(k)\|\leq\eta.
\end{equation}
\end{definition}

\begin{lemma}
\label{lem:fixed_cluster}
Suppose $\Pcal(k)=\Pcal$ for all $k\in\mathbb{Z}_{\geq 0}$ and let $\Ccal\in\Pcal$.
Under Assumption~\ref{ass:tube_ingredients},
problem~\eqref{eq:nom_opt_C} is recursively feasible,
\begin{equation}
J_\Ccal^\star(x_\Ccal(k))\longrightarrow0,
\label{eq:fixed_Jstar_convergence}
\end{equation}
and
\begin{equation}
\limsup_{k\to\infty}\|x_\Ccal(k)\|
\leq
\rho_\Ccal,
\qquad
\rho_\Ccal:=
\sqrt{
\frac{\Gamma_\Ccal}{\lambda_{\min}(Q_\Ccal)}
}.
\label{eq:fixed_cluster_bound}
\end{equation}
\end{lemma}

\begin{proof}
Let
$(\mathbf z_\Ccal^\star(k),\mathbf v_\Ccal^\star(k))$
be optimal at time $k$. By~\eqref{eq:shifted_candidate}, with the
indices shifted from $k$ to $k+1$, a feasible candidate exists at the
successor state. Hence recursive feasibility follows. From
\eqref{eq:shifted_candidate_decrease} and optimality at time $k+1$,
\begin{equation}
J_\Ccal^\star(x_\Ccal(k+1))
-
J_\Ccal^\star(x_\Ccal(k))
\leq
-
\ell_\Ccal
\left(
z^\star_{\Ccal,0|k},
v^\star_{\Ccal,0|k}
\right).
\label{eq:fixed_Jstar_decrease}
\end{equation}
Since $J_\Ccal^\star\geq0$, this implies
\[
\sum_{k=0}^{\infty}
\ell_\Ccal
\left(
z^\star_{\Ccal,0|k},
v^\star_{\Ccal,0|k}
\right)
<\infty.
\]
Because $Q_\Ccal\succ0$,
\[
\ell_\Ccal
\left(
z^\star_{\Ccal,0|k},
v^\star_{\Ccal,0|k}
\right)
\geq
\lambda_{\min}(Q_\Ccal)
\|z^\star_{\Ccal,0|k}\|^2,
\]
and therefore
\begin{equation}
z^\star_{\Ccal,0|k}\longrightarrow0.
\label{eq:fixed_z0_convergence}
\end{equation}

Since
$0\in\operatorname{int}(\Zcal_{\mathrm f,\Ccal})$, for sufficiently
large $k$ the terminal controller is feasible from
$z^\star_{\Ccal,0|k}$. Repeated application of
\eqref{eq:terminal_lyapunov_decrease} therefore gives
\[
0
\leq
J_\Ccal^\star(x_\Ccal(k))
\leq
V_{\mathrm f,\Ccal}
\left(z^\star_{\Ccal,0|k}\right),
\]
which together with~\eqref{eq:fixed_z0_convergence} proves
\eqref{eq:fixed_Jstar_convergence}.

Finally, using the optimal nominal sequence itself as the candidate
in~\eqref{eq:upper_bound},
\[
J_\Ccal^{\mathrm r}
\left(
\mathbf x_\Ccal^\star(k),
\mathbf u_\Ccal^\star(k)
\right)
\leq
\left(
\sqrt{J_\Ccal^\star(x_\Ccal(k))}
+
\sqrt{\Gamma_\Ccal}
\right)^2.
\]
Hence
\[
\limsup_{k\to\infty}
J_\Ccal^{\mathrm r}
\left(
\mathbf x_\Ccal^\star(k),
\mathbf u_\Ccal^\star(k)
\right)
\leq
\Gamma_\Ccal.
\]
Since
$x^\star_{\Ccal,0|k}=x_\Ccal(k)$,
\[
J_\Ccal^{\mathrm r}
\left(
\mathbf x_\Ccal^\star(k),
\mathbf u_\Ccal^\star(k)
\right)
\geq
\lambda_{\min}(Q_\Ccal)\|x_\Ccal(k)\|^2,
\]
which proves~\eqref{eq:fixed_cluster_bound}.
\end{proof}

For a fixed partition, Lemma~\ref{lem:fixed_cluster} directly gives
\begin{equation}
\limsup_{k\to\infty}\|x(k)\|
\leq
\left(
\sum_{\Ccal\in\Pcal}\rho_\Ccal^2
\right)^{1/2}.
\label{eq:fixed_partition_global_bound}
\end{equation}
Alternatively, the tube relation and
\eqref{eq:fixed_z0_convergence} give
\[
\limsup_{k\to\infty}\|x_\Ccal(k)\|
\leq
r_\Ccal,
\qquad
r_\Ccal:=
\max_{e_\Ccal\in\Ecal_\Ccal}\|e_\Ccal\|,
\]
so either bound may be used.

\subsubsection{Predicted performance} We next describe the predicted performance of trajectories generated by the algorithm, relative to centralized MPC.

\begin{theorem}
\label{thm:performance}
At every time $k\in\mathbb{Z}_{\geq 0}$, let
$\mathbf x^{\mathrm{alg}}(k),\mathbf u^{\mathrm{alg}}(k)$ and
$\mathbf x^{\mathrm{cen}}(k),\mathbf u^{\mathrm{cen}}(k)$ denote the
real predicted trajectories selected by Algorithm~\ref{alg:dyn} and centralized MPC respectively, both
initialized at $x(k)$.

\begin{align}
\sum_{t=0}^{N_{\mathrm p}-1}
\ell
\left(
x^{\mathrm{alg}}_{t|k},
u^{\mathrm{alg}}_{t|k}
\right)
\leq&
\sum_{t=0}^{N_{\mathrm p}-1}
\ell
\left(
x^{\mathrm{cen}}_{t|k},
u^{\mathrm{cen}}_{t|k}
\right)
\nonumber\\
&+
\sum_{\Ccal\in\Pcal(k)}
\delta_\Ccal(x_\Ccal(k)).
\label{eq:performance_bound}
\end{align}
\end{theorem}

\begin{proof}
Using~\eqref{eq:upper_bound}, the mismatch definition, and
Lemma~\ref{lem:local_lower_bound},
\begin{align}
\sum_{t=0}^{N_{\mathrm p}-1}
\ell
\left(
x^{\mathrm{alg}}_{t|k},
u^{\mathrm{alg}}_{t|k}
\right)
&\leq
\sum_{\Ccal\in\Pcal(k)}
\bar J_\Ccal
\left(
\tilde{\mathbf z}_\Ccal(k),
\tilde{\mathbf v}_\Ccal(k)
\right)
\nonumber\\
&=
\sum_{\Ccal\in\Pcal(k)}
\underline J_\Ccal(x_\Ccal(k))
+
\sum_{\Ccal\in\Pcal(k)}
\delta_\Ccal(x_\Ccal(k))
\nonumber\\
&\leq
\sum_{t=0}^{N_{\mathrm p}-1}
\ell
\left(
x^{\mathrm{cen}}_{t|k},
u^{\mathrm{cen}}_{t|k}
\right)
+
\sum_{\Ccal\in\Pcal(k)}
\delta_\Ccal(x_\Ccal(k)),
\end{align}
which proves the result.
\end{proof}

\subsubsection{Recursive feasibility}

\begin{definition}
A partition $\Pcal$ is \emph{feasible} at $x$ if
problem~\eqref{eq:nom_opt_C} is feasible for every
$\Ccal\in\Pcal$ at the cluster state $x_\Ccal$.
\end{definition}

\begin{assumption}
\label{ass:initial_partition}
The initial partition
$\Pcal_{\mathrm{ini}}\in\Pcal_{\mathrm{adm}}$
is feasible at $x(0)$ and a feasible candidate for
problem~\eqref{eq:nom_opt_C} is available for every
$\Ccal\in\Pcal_{\mathrm{ini}}$.
\end{assumption}

\begin{theorem}
\label{thm:recursive_feasibility}
Suppose Assumptions~\ref{ass:tube_ingredients} and
\ref{ass:initial_partition} hold and
$\Pcal_{\mathrm{adm}}$ is finite. Then Algorithm~\ref{alg:dyn}
is recursively feasible: at every $k\in\mathbb{Z}_{\geq 0}$ the partition
selection~\eqref{eq:partition-selection} admits a finite-valued
minimizer and problem~\eqref{eq:nom_opt_C} is feasible for every
$\Ccal\in\Pcal(k)$.
\end{theorem}

\begin{proof}
At $k=0$, Assumption~\ref{ass:initial_partition} provides the
finite-valued candidate partition $\Pcal_{\mathrm{ini}}$. Hence, since
$\Pcal_{\mathrm{adm}}$ is finite,
\eqref{eq:partition-selection} admits a finite-valued minimizer
$\Pcal(0)$, whose clusters necessarily admit feasible candidates.

Suppose now that $\Pcal(k)$ has been selected and its cluster MPC
problems solved. By~\eqref{eq:shifted_candidate}, every
$\Ccal\in\Pcal(k)$ has a feasible shifted candidate at $x_\Ccal(k+1)$.
Thus the previous partition $\Pcal(k)$ has finite aggregate value at
time $k+1$ and is a feasible competitor in
\eqref{eq:partition-selection}. Consequently a finite-valued minimizer
$\Pcal(k+1)$ exists. Since Algorithm~\ref{alg:dyn} assigns infinite
value whenever no feasible candidate is available, every cluster of
$\Pcal(k+1)$ is feasible for~\eqref{eq:nom_opt_C}. The result follows
by induction.
\end{proof}

\subsubsection{Practical stability under switching}

For an admissible partition $\Pcal$, define
\begin{equation}
\mathcal G_\Pcal(k)
:=
\sum_{\Ccal\in\Pcal}
\left[
\bar J_\Ccal
\left(
\tilde{\mathbf z}_\Ccal(k),
\tilde{\mathbf v}_\Ccal(k)
\right)
+
f_{\mathrm{com}}(\Ccal)
\right],
\label{eq:partition_G}
\end{equation}
with $\mathcal G_\Pcal(k)=\infty$ if any cluster has no feasible
candidate. Since
\begin{equation}
\sum_{\Ccal\in\Pcal}
\underline J_\Ccal(x_\Ccal(k))
=
\sum_{i=1}^{N}J_i^{\min}(x_i(k))
=:L(x(k))
\label{eq:partition_independent_lower}
\end{equation}
is independent of $\Pcal$, the partition rule
\eqref{eq:partition-selection} is equivalently
\begin{equation}
\Pcal(k)
\in
\arg\min_{\Pcal\in\Pcal_{\mathrm{adm}}}
\mathcal G_\Pcal(k).
\label{eq:G_partition_minimizer}
\end{equation}

\begin{theorem}
\label{thm:switched_practical_stability}
Suppose Assumptions~\ref{ass:tube_ingredients} and
\ref{ass:initial_partition} hold,
$\Pcal_{\mathrm{adm}}$ is finite, and
$f_{\mathrm{com}}(\Ccal)\geq0$ for every
$\Ccal\in\Ccal_{\mathrm{adm}}$. Define
\begin{equation}
r_{\max}
:=
\max_{\Pcal\in\Pcal_{\mathrm{adm}}}
\;
\max_{
e\in\prod_{\Ccal\in\Pcal}\Ecal_\Ccal
}
\|e\|.
\label{eq:rmax}
\end{equation}
Then every closed-loop trajectory generated by
Algorithm~\ref{alg:dyn} satisfies
\begin{equation}
\limsup_{k\to\infty}\|x(k)\|
\leq
r_{\max}.
\label{eq:switched_tube_practical_bound}
\end{equation}
Hence the switched closed-loop system is practically asymptotically
stable.
\end{theorem}

\begin{proof}
Recursive feasibility follows from
Theorem~\ref{thm:recursive_feasibility}. For
$\Ccal\in\Pcal(k)$, let
$(\tilde{\mathbf z}_\Ccal(k),\tilde{\mathbf v}_\Ccal(k))$
be the candidate used in the partition evaluation and
$(\mathbf z_\Ccal^\star(k),\mathbf v_\Ccal^\star(k))$
the subsequently obtained optimal solution. Then
\begin{equation}
J_\Ccal^\star(x_\Ccal(k))
\leq
\Phi_\Ccal
\left(
\tilde{\mathbf z}_\Ccal(k),
\tilde{\mathbf v}_\Ccal(k)
\right).
\label{eq:switched_optimal_candidate}
\end{equation}
Combining this with~\eqref{eq:shifted_candidate_decrease} gives
\begin{align}
\Phi_\Ccal
\left(
\hat{\mathbf z}_\Ccal(k+1),
\hat{\mathbf v}_\Ccal(k+1)
\right)
\leq&
\Phi_\Ccal
\left(
\tilde{\mathbf z}_\Ccal(k),
\tilde{\mathbf v}_\Ccal(k)
\right)
\nonumber\\
&-
\ell_\Ccal
\left(
z^\star_{\Ccal,0|k},
v^\star_{\Ccal,0|k}
\right).
\label{eq:switched_Phi_decrease}
\end{align}

From~\eqref{eq:upper_bound_problem},
\(
\bar J_\Ccal=h_\Ccal(\Phi_\Ccal), 
h_\Ccal(s):=
\left(\sqrt{s}+\sqrt{\Gamma_\Ccal}\right)^2.
\)
For $s_1\geq s_2\geq0$,
\(
h_\Ccal(s_1)-h_\Ccal(s_2)\geq s_1-s_2,
\)
and therefore~\eqref{eq:switched_Phi_decrease} implies
\begin{align}
\bar J_\Ccal
\left(
\hat{\mathbf z}_\Ccal(k+1),
\hat{\mathbf v}_\Ccal(k+1)
\right)
\leq&
\bar J_\Ccal
\left(
\tilde{\mathbf z}_\Ccal(k),
\tilde{\mathbf v}_\Ccal(k)
\right)
\nonumber\\
&-
\ell_\Ccal
\left(
z^\star_{\Ccal,0|k},
v^\star_{\Ccal,0|k}
\right).
\label{eq:switched_barJ_decrease}
\end{align}
Since the shifted candidates are used at time $k+1$ for all clusters
of the previous partition,
\begin{equation}
\mathcal G_{\Pcal(k)}(k+1)
\leq
\mathcal G_{\Pcal(k)}(k)
-
\sum_{\Ccal\in\Pcal(k)}
\ell_\Ccal
\left(
z^\star_{\Ccal,0|k},
v^\star_{\Ccal,0|k}
\right).
\label{eq:old_partition_G_decrease}
\end{equation}
The old partition is feasible at $k+1$ by
\eqref{eq:shifted_candidate}, while
\eqref{eq:G_partition_minimizer} gives
\begin{equation}
\mathcal G_{\Pcal(k+1)}(k+1)
\leq
\mathcal G_{\Pcal(k)}(k+1).
\label{eq:new_partition_no_larger}
\end{equation}
Consequently, with
$\mathcal G(k):=\mathcal G_{\Pcal(k)}(k)$,
\begin{equation}
\mathcal G(k+1)-\mathcal G(k)
\leq
-
\sum_{\Ccal\in\Pcal(k)}
\ell_\Ccal
\left(
z^\star_{\Ccal,0|k},
v^\star_{\Ccal,0|k}
\right).
\label{eq:switched_G_decrease}
\end{equation}
Because $\bar J_\Ccal\geq0$ and
$f_{\mathrm{com}}(\Ccal)\geq0$, $\mathcal G(k)$ is bounded below.
Summing~\eqref{eq:switched_G_decrease} therefore gives
\begin{equation}
\sum_{k=0}^{\infty}
\sum_{\Ccal\in\Pcal(k)}
\ell_\Ccal
\left(
z^\star_{\Ccal,0|k},
v^\star_{\Ccal,0|k}
\right)
<\infty.
\label{eq:switched_stage_summability}
\end{equation}

Let
\[
q_{\min}:=
\min_{i\in\Ncal}\lambda_{\min}(Q_i)>0.
\]
Since the active clusters partition the agent set,
\[
\sum_{\Ccal\in\Pcal(k)}
\ell_\Ccal
\left(
z^\star_{\Ccal,0|k},
v^\star_{\Ccal,0|k}
\right)
\geq
q_{\min}\|z^\star_{0|k}\|^2.
\]
Thus~\eqref{eq:switched_stage_summability} implies
$z^\star_{0|k}\to0$. Finally,
$x(k)=z^\star_{0|k}+e(k)$ with
$e(k)\in\prod_{\Ccal\in\Pcal(k)}\Ecal_\Ccal$, and hence
\[
\limsup_{k\to\infty}\|x(k)\|
\leq
r_{\max},
\]
which proves the result.
\end{proof}

\begin{corollary}
\label{cor:switched_real_cost_bound}
Under the conditions of
Theorem~\ref{thm:switched_practical_stability}, define
\[
q_{\min}:=
\min_{i\in\Ncal}\lambda_{\min}(Q_i),
\qquad
\Gamma_{\max}
:=
\max_{\Pcal\in\Pcal_{\mathrm{adm}}}
\sum_{\Ccal\in\Pcal}\Gamma_\Ccal.
\]
Then
\begin{equation}
\limsup_{k\to\infty}\|x(k)\|
\leq
\min\left\{
r_{\max},
\sqrt{\frac{\Gamma_{\max}}{q_{\min}}}
\right\}.
\label{eq:switched_real_cost_state_bound}
\end{equation}
\end{corollary}

\begin{proof}
Since $z^\star_{0|k}\to0$ and
$0\in\operatorname{int}(\Zcal_{\mathrm f,\Ccal})$ for every admissible
cluster, the terminal controller is eventually feasible from every
active $z^\star_{\Ccal,0|k}$. Hence
$J_\Ccal^\star(x_\Ccal(k))\to0$ uniformly over the finite admissible
cluster family. Using the optimal nominal trajectory in
\eqref{eq:upper_bound} therefore gives
\[
\limsup_{k\to\infty}
\sum_{\Ccal\in\Pcal(k)}
J_\Ccal^{\mathrm r}
\left(
\mathbf x_\Ccal^\star(k),
\mathbf u_\Ccal^\star(k)
\right)
\leq
\Gamma_{\max}.
\]
Since the real predictions are initialized at $x(k)$,
\[
\sum_{\Ccal\in\Pcal(k)}
J_\Ccal^{\mathrm r}
\left(
\mathbf x_\Ccal^\star(k),
\mathbf u_\Ccal^\star(k)
\right)
\geq
q_{\min}\|x(k)\|^2.
\]
Combining both inequalities gives
$\limsup_{k\to\infty}\|x(k)\|
\leq\sqrt{\Gamma_{\max}/q_{\min}}$, which together with
\eqref{eq:switched_tube_practical_bound} proves the result.
\end{proof}

\section{Simulation Results}\label{sec:system}
\subsection{Physical example and implementation}
We consider the eight horizontally guided masses in
Fig.~\ref{fig:ring-graph-clusters}, with $x_i=\col(q_i,\dot q_i)$ and
horizontal force input $u_i$. Each edge is an unprestressed axial
spring--damper pair; each mass also has a spring and damper to ground.
Table~\ref{tab:plants} gives the masses $m_i$, ground coefficients
$k_i^{\mathrm g},d_i^{\mathrm g}$, base axial coefficients
$\kappa_{ij},c_{ij}$ and projection factors
$p_{ij}=\cos^2\theta_{ij}$. For plant scale $s$, define
$k_{ij}=s p_{ij}\kappa_{ij}$ and $d_{ij}=s p_{ij}c_{ij}$.
These coefficients are symmetric and zero for absent edges. Let
$k_i^\Sigma=k_i^{\mathrm g}+\sum_{j\in\Ncal_i}k_{ij}$ and
$d_i^\Sigma=d_i^{\mathrm g}+\sum_{j\in\Ncal_i}d_{ij}$.
Forward Euler with $T_{\mathrm s}=0.1\,\mathrm{s}$ gives the blocks
in \eqref{eq:model} explicitly:
\begin{equation}
\begin{aligned}
A_{ii}&=\begin{bmatrix}
1&T_{\mathrm s}\\
-T_{\mathrm s}k_i^\Sigma/m_i&1-T_{\mathrm s}d_i^\Sigma/m_i
\end{bmatrix},\\
A_{ij}&=\begin{bmatrix}
0&0\\T_{\mathrm s}k_{ij}/m_i&T_{\mathrm s}d_{ij}/m_i
\end{bmatrix}\quad (j\ne i),\\
B_{ii}&=\begin{bmatrix}0\\T_{\mathrm s}/m_i\end{bmatrix},
\qquad B_{ij}=\begin{bmatrix}0\\0\end{bmatrix}\quad(j\ne i).
\end{aligned}
\label{eq:simulation_blocks}
\end{equation}
Thus coupling is through states only; unequal masses generally give
$A_{ij}\ne A_{ji}$ even though the mechanical graph is undirected.

\begin{table*}[t]
\centering
\caption{Plant parameters and common initial condition (SI units). 
For each edge, $k_{ij}=s p_{ij}\kappa_{ij}$ and $d_{ij}=s p_{ij}c_{ij}$. 
Equilibrium coordinates are given in Fig.~\ref{fig:ring-graph-clusters}.}
\label{tab:plants}
\scriptsize
\setlength{\tabcolsep}{3pt}

\begin{minipage}[t]{0.43\textwidth}
\centering
\begin{tabular}{c|rrrrr}
\hline
$i$ & $m_i$ & $k_i^{\rm g}$ & $d_i^{\rm g}$ & $q_i(0)$ & $\dot q_i(0)$\\
\hline
1 & 1.00 & 0.80 & 2.80 &  0.049 & $-0.199$\\
2 & 1.15 & 1.00 & 3.00 & $-0.044$ &  0.160\\
3 & 0.85 & 0.90 & 2.60 &  0.042 & $-0.170$\\
4 & 1.30 & 0.75 & 3.20 & $-0.043$ &  0.150\\
5 & 0.95 & 1.10 & 2.70 &  0.041 & $-0.160$\\
6 & 1.10 & 0.85 & 2.90 & $-0.042$ &  0.170\\
7 & 0.90 & 1.00 & 2.50 &  0.044 & $-0.150$\\
8 & 1.25 & 0.95 & 3.10 & $-0.041$ &  0.160\\
\hline
\end{tabular}
\end{minipage}
\hfill
\begin{minipage}[t]{0.31\textwidth}
\centering
\begin{tabular}{c|rrr}
\hline
$(i,j)$ & $\kappa_{ij}$ & $c_{ij}$ & $p_{ij}$\\
\hline
$(1,2)$ & 1.00  & 0.70  & 0.5\\
$(2,3)$ & 0.70  & 1.10  & 0.5\\
$(2,4)$ & 0.104 & 0.064 & 0.5\\
$(4,5)$ & 0.90  & 1.20  & 0.5\\
$(4,6)$ & 1.15  & 0.90  & 0.5\\
$(6,7)$ & 0.80  & 0.60  & 0.5\\
$(7,8)$ & 0.50  & 0.40  & 0.5\\
$(1,3)$ & 0.80  & 0.70  & 1.0\\
$(5,7)$ & 0.90  & 0.80  & 0.5\\
\hline
\end{tabular}
\end{minipage}
\hfill
\begin{minipage}[t]{0.21\textwidth}
\centering
\begin{tabular}{c|rr}
\hline
Plant & $s$ & $\lambda$\\
\hline
1 & 0.12  & 0.0012\\
2 & 0.012 & 0.00014\\
3 & 0.005 & 0.00035\\
\hline
\end{tabular}
\end{minipage}
\end{table*}

The constraints \eqref{eq:constraints_local} impose
$|q_i|\leq0.05\,\mathrm{m}$, $|\dot q_i|\leq0.2\,\mathrm{m/s}$ and
$|u_i|\leq2\,\mathrm{N}$. We use $Q_i=\diag(1,0.1)$, $R_i=0.2$,
$N_{\mathrm p}=20$ and 30 closed-loop samples. All 68 nonempty connected
clusters are admissible, inducing 240 partitions, including centralized
and decentralized operation. For each cluster, discrete LQR gains,
Trodden's LP tube~\cite{trodden2016one}, tightened constraints and terminal
ingredients are computed offline as in Section~\ref{sec:tube_mpc}.
The communication penalty is \eqref{eq:communication-penalty},
with $\phi(\Ccal)$ counting internal graph edges and the fixed prices
in Table~\ref{tab:plants}.

Algorithm~\ref{alg:dyn} starts centralized and uses the exhaustive
selection \eqref{eq:partition-selection}. Feasible shifted candidates
are retained; otherwise the cluster QP supplies a candidate.
No partition schedule or monotonic splitting rule is imposed.
The centralized, robust decentralized and adaptive controllers use
the same initial state and propagate their own trajectories through
the full coupled discrete model. The common initial state has no zero coordinates.
All offline certificates and online constraint and decrease checks pass.

\subsection{Cooperation and realized performance}
\begin{figure}[t]
\centering
\includegraphics[width=\columnwidth]{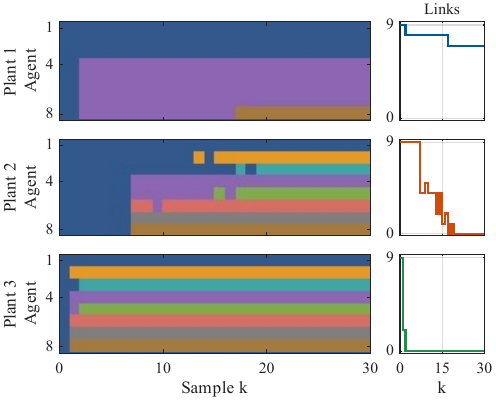}
\caption{Adaptive partitions (left) and internal communication links
(right). Equal colours within a column identify a cluster; singleton
agents retain distinct colours. All panels use the same 30-sample
time window. Recoupling is permitted by the selection rule.}
\label{fig:partition_evolution}
\end{figure}

Figure~\ref{fig:partition_evolution} shows distinct cooperation patterns.
Plant~1 splits at $k=2$ into $\{1,2,3\}$ and $\{4,5,6,7,8\}$, retaining
these large groups until agent~8 separates at $k=17$. It remains
coupled as its state approaches the origin. Its robust decentralized
QP is infeasible at $k=0$; this outcome is recorded rather than
replaced by a fallback controller. In contrast, decentralized operation
is initially feasible for Plants~2 and~3.
Plant~2 first splits at $k=7$ and becomes decentralized at $k=17$.
A brief recoupling at $k=18$ precedes sustained decentralization from
$k=19$. Plant~3 reaches singleton operation at $k=2$.
The right-hand panels show internal links, rather than differently
scaled penalties, so communication effort can be compared directly.
Since both $s$ and $\lambda$ vary, these examples illustrate tuned
trade-offs, not an isolated study of coupling strength.

\begin{table}[t]
\centering
\caption{Realized closed-loop stage cost over $k=0,\ldots,29$.
Infeasibility refers to the decentralized QP at $k=0$.}
\label{tab:performance_comparison}
\setlength{\tabcolsep}{3pt}\small
\begin{tabular}{lrrr}
\hline
MPC architecture & Plant 1 & Plant 2 & Plant 3\\
\hline
Centralized & 0.08928796 & 0.09377568 & 0.09409090\\
Robust decentralized & Infeasible & 0.09377557 & 0.09409083\\
Adaptive clustering & 0.08928776 & 0.09377530 & 0.09409082\\
\hline
\end{tabular}
\end{table}

Table~\ref{tab:performance_comparison} sums the realized stage cost
$\ell(x(k),u(k))$ along each controller's own trajectory; terminal
and communication penalties are excluded. The feasible controllers
have nearly identical costs. The slightly smaller adaptive values
are not a claim of superiority to centralized finite-horizon MPC:
the independent receding-horizon trajectories differ. The main
distinction is the cooperation pattern and, for Plant~1, the ability
to initialize when robust decentralized operation is infeasible.

\begin{figure}[t]
\centering
\includegraphics[width=\columnwidth]{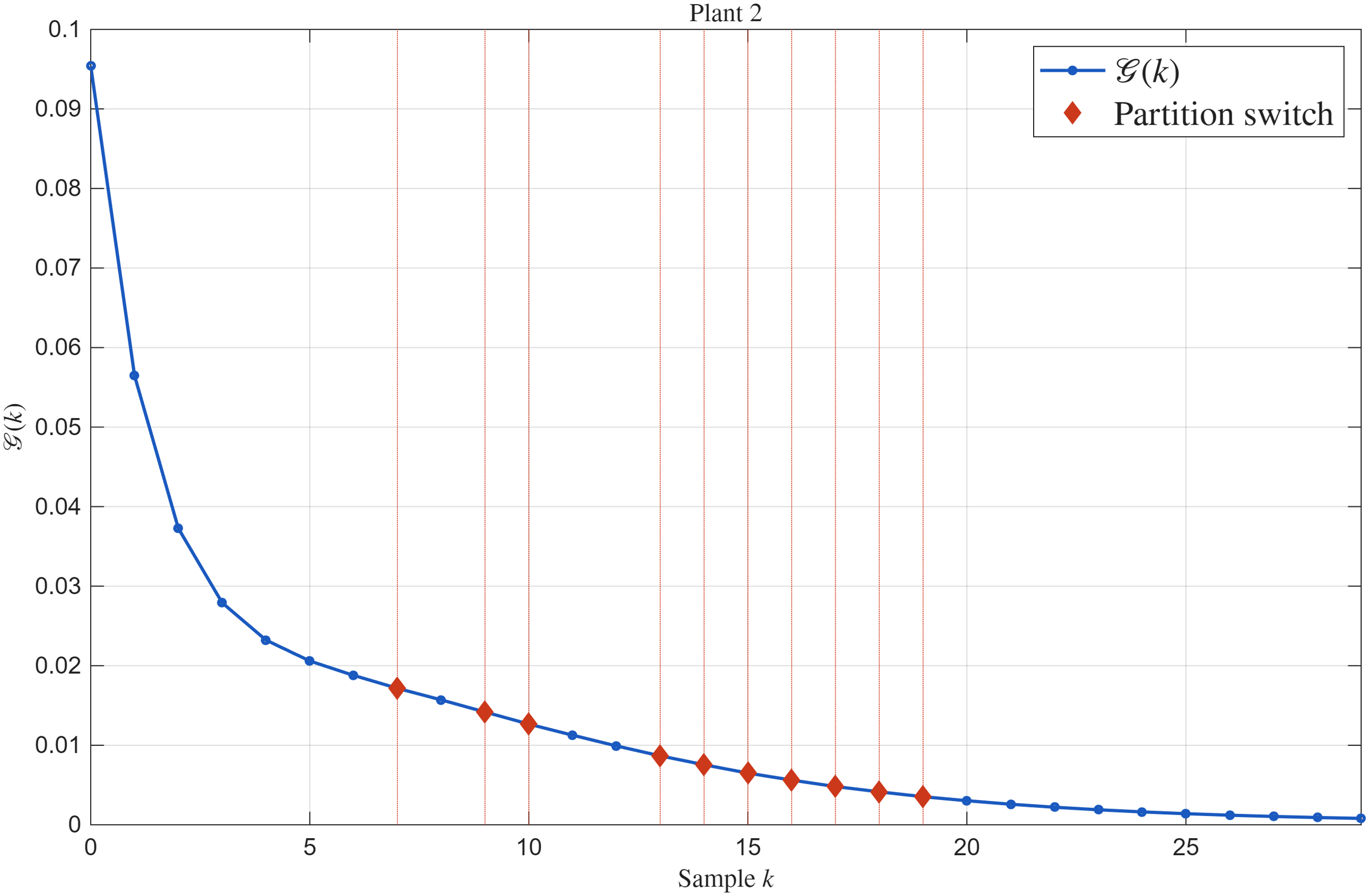}
\caption{Simulation of the Lyapunov candidate $\Gcal(k)$ and its descent \eqref{eq:switched_G_decrease}, as outlined in the proof of Theorem~\ref{thm:switched_practical_stability}, for Plant~2.}
\label{fig:sim_lyapunov}
\end{figure}

FInally, Figure~\ref{fig:sim_lyapunov} displays the Lyapunov candidate $\Gcal(k)$, and highlights the partition switches. It can be seen that despite the switched behavior of the dynamics at the switching times, the function remains non-increasing.

\subsection{Mismatch bounds and local interpretation}
\begin{figure}[t]
\centering
\includegraphics[width=\columnwidth]{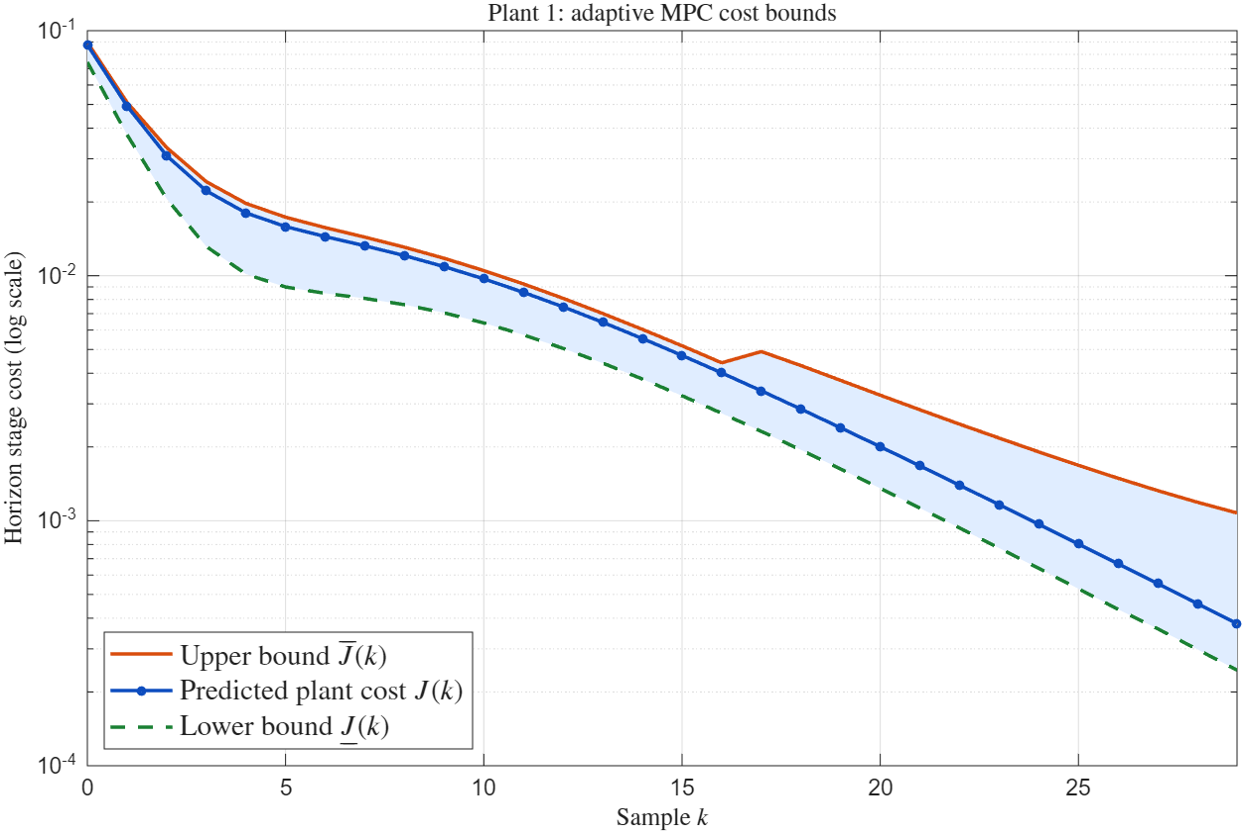}
\caption{Plant~1: predicted full-system cost $J_{\mathrm{alg}}$,
bracketed by the aggregate optimistic lower and robust candidate
upper bounds. The shaded band spans these bounds. All values use the adaptive state $x(k)$ and selected
candidates. The cost axis is logarithmic.}
\label{fig:sim_upper_lower}
\end{figure}

\begin{figure}[t]
\centering
\includegraphics[width=\columnwidth]{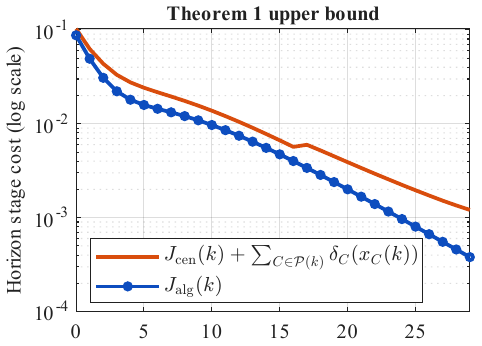}
\caption{Plant~1, Theorem~\ref{thm:performance}:
the predicted algorithmic horizon stage cost $J_{\mathrm{alg}}$ and
the certified bound $J_{\mathrm{cen}}+\Delta$ on a logarithmic cost
axis. Both predictions start at the same adaptive state; the
inequality holds at all 30 displayed samples.}
\label{fig:sim_thm1}
\end{figure}

At each adaptive state, we also evaluate the two $N_{\mathrm p}$-step
stage costs in Theorem~\ref{thm:performance}, keeping the selected
partition fixed during prediction and propagating the full dynamics.
These same-state predictions differ from the closed-loop totals in
Table~\ref{tab:performance_comparison}.
Figure~\ref{fig:sim_upper_lower} brackets the predicted full-system cost by
the aggregate lower and upper bounds. Figure~\ref{fig:sim_thm1} compares
$J_{\mathrm{alg}}$ with the bound $J_{\mathrm{cen}}+\Delta$, where
$\Delta(k):=\sum_{\Ccal\in\Pcal(k)}\delta_\Ccal(x_\Ccal(k))$.
Both predictions start at the same adaptive state. All 90 tested inequalities hold.
The largest positive predicted excess is $1.81\times10^{-7}$, while the
maximum mismatch bounds for Plants~1--3 are respectively
$0.01523$, $0.00576$ and $0.00562$. The bound is therefore conservative;
a small observed gap does not imply a tight certificate.

\begin{figure}[t]
\centering
\includegraphics[width=\columnwidth]{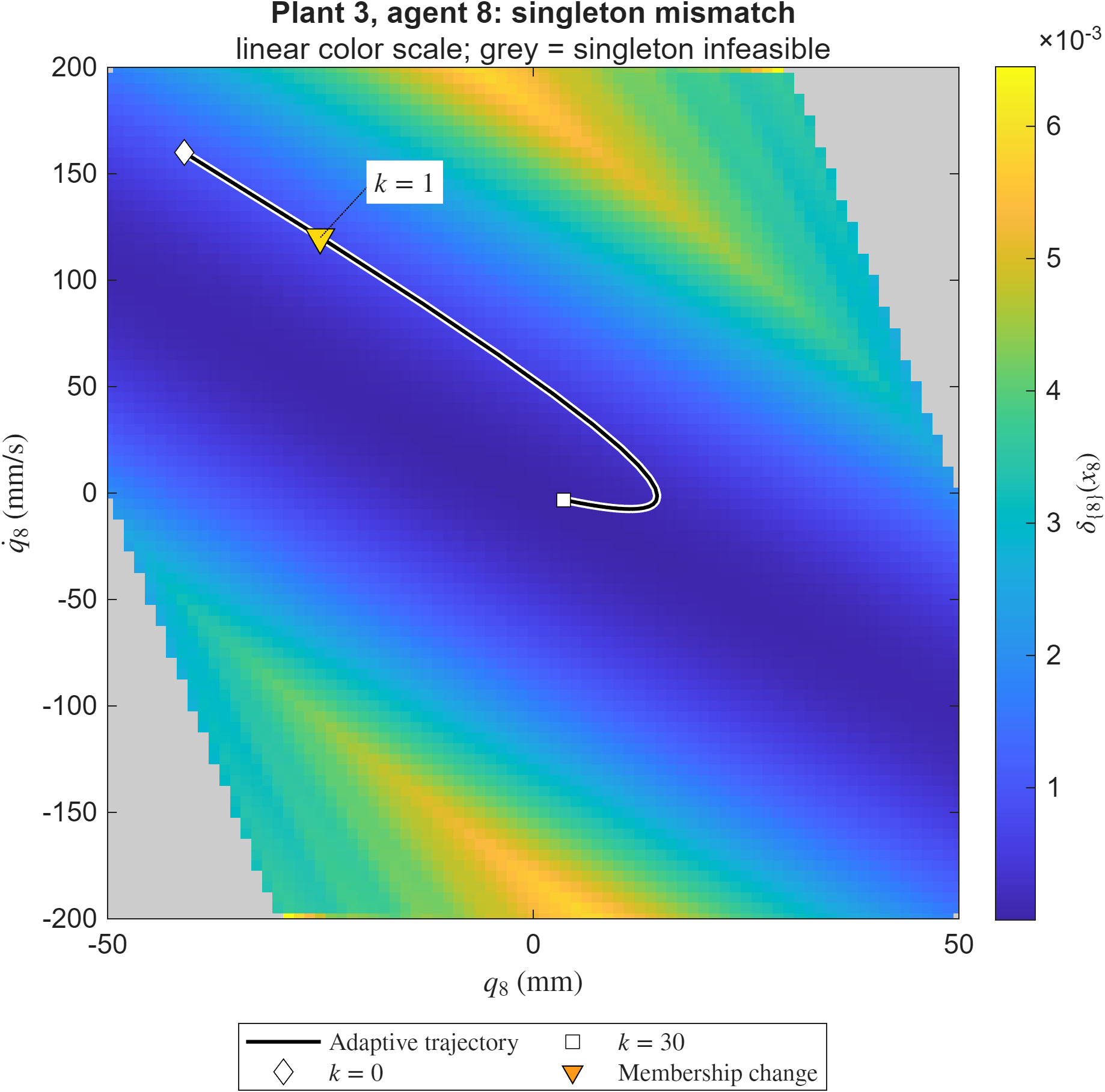}
\caption{Plant~3, agent~8: freshly optimized singleton mismatch and
adaptive trajectory (black). Grey denotes singleton infeasibility;
the inset resolves the initial state (diamond) near its boundary.
The triangle marks the membership change at $k=2$, and the square
marks $k=30$. Axis coordinates also show the local state limits.}
\label{fig:sim_delta}
\end{figure}

Finally, Fig.~\ref{fig:sim_delta} evaluates the singleton mismatch
\eqref{eq:deltaC} for agent~8 of Plant~3 using freshly optimized candidates
at each plotted state. The black curve is the agent's adaptive trajectory.
The highlighted marker at $k=2$ records its transition from the
centralized cluster to singleton operation. Grey denotes singleton
infeasibility, not infeasibility of the adaptive controller.
The map uses fresh singleton solutions, so its values need not equal
the mismatch of a retained online candidate.

\section{Conclusion}
\label{sec:conclusion}

This paper developed a dynamic clustering framework for tube-based MPC of constrained, dynamically coupled multi-agent systems. Offline cluster ingredients support online partition selection through a certified mismatch measure and a fixed communication penalty. Under the stated assumptions, the resulting switching scheme preserves recursive feasibility, constraint satisfaction, and practical stability. Theorem~\ref{thm:performance} additionally bounds the predicted finite-horizon stage-cost difference relative to centralized MPC, with both predictions initialized at the same state.

The eight-agent mechanical example illustrates how this framework accommodates different degrees of cooperation. Across the three plant configurations, the adaptive controller achieved realized closed-loop costs close to those of centralized MPC while reducing communication. For the strongly coupled plant, it remained feasible from an initial state outside the decentralized controller's feasible set and retained cooperation throughout the simulation. The other configurations reached decentralized operation at different rates, with temporary recoupling in the intermediate case. These results illustrate that partition adaptation need not be monotonic. The numerical checks also confirmed the predicted mismatch bound, although its conservatism leaves room for improvement.

Future work will focus on tighter descriptions of inter-cluster disturbances and scalable partition selection. Reducing the conservatism of the mismatch certificate while retaining the feasibility and descent conditions could improve the trade-off between communication and control performance in larger networks.

\bibliographystyle{IEEEtran}
\bibliography{biblio}

%
\IEEEpeerreviewmaketitle

%





\end{document}